\newcommand{\trm}[1]{\textrm{#1}}
\newcommand{\prens}[1]{\left(#1\right)}
\newcommand{\perm}{\varepsilon}
\renewcommand{\u}{\mathbf{u}}
\newcommand{\E}{\mathbf{e}}
\newcommand{\um}{\mu\trm{m}}
\newcommand{\TE}[1]{TE\(_#1\)}

\newcommand{\tbmu}{t_{\trm{b}\mu}}

\documentclass[reprint,amsmath,amssymb,aps]{revtex4-1}

\usepackage{graphicx}
\usepackage{xcolor}
\usepackage{tablefootnote}

\begin{document}

\title[Acousto-optic modulation of a wavelength-scale waveguide]{Acousto-optic modulation of a wavelength-scale waveguide}

\author{Christopher J. Sarabalis}
\email{sicamor@stanford.edu}
\author{Rapha\"{e}l Van Laer}
\author{Rishi N. Patel}
\author{Yanni D. Dahmani}
\author{Wentao Jiang}
\author{Felix M. Mayor}
\author{Amir H. Safavi-Naeini}
\email{safavi@stanford.edu}

\affiliation{%
 Department of Applied Physics and Ginzton Laboratory, Stanford University\\
 348 Via Pueblo Mall, Stanford, California 94305, USA
}%

\date{\today}%

\begin{abstract}
	We demonstrate a collinear acousto-optic modulator in a suspended film of lithium niobate employing a high-confinement, wavelength-scale waveguide. By strongly confining the optical and mechanical waves, this modulator improves by orders of magnitude a figure-of-merit that accounts for both acousto-optic and electro-mechanical efficiency. Our device demonstration marks a significant technological advance in acousto-optics that promises a novel class of compact and low-power frequency shifters, tunable filters, non-magnetic isolators, and beam deflectors.
\end{abstract}

\maketitle

\section{Introduction}

Lithium niobate (LN) has played a central role in the development of acousto-optic (AO) devices. In the decades following the initial demonstration of the AO tunable filter~\cite{Harris1970}, the electrical power consumption of these devices improved from watts to milliwatts through the development of Ti-indiffused and proton-exchange LN optical and surface acoustic wave (SAW) waveguides, and through the development of efficient SAW transducers~\cite{Ohmachi1977,Binh1980,Smith1990,Hinkov1994,Duchet1995}. These waveguides only weakly confine the optical and mechanical fields. The potential for greater confinement and thereby larger interaction strengths is an opportunity to again dramatically improve the efficiency and/or reduce the size of these devices.

In recent years a new LN waveguide technology has emerged that can vastly improve electro-optic, acousto-optic, and nonlinear optical devices. Advances in etch techniques have resulted in low-loss, wavelength-scale optical waveguides~\cite{Wang2014} in high-quality, single-crystal films of LN~\cite{Levy1998}. Owing to their high confinement, these waveguides have powered the development of an array of compact, highly efficient nonlinear devices~\cite{Chang2016, Rao2016, Yu2019} and modulators~\cite{Wang2018, Mckenna2020, Sarabalis2020} for classical and quantum applications. In tandem, LN films have been used to realize low-loss, strongly-coupled piezoelectric devices~\cite{Olsson2014,Vidal2017,Pop2017,Manzaneque2019,Sarabalis2020sband,Mayor2020}. We have recently demonstrated that high-confinement waveguides in this platform can be efficiently piezoelectrically transduced~\cite{Dahmani2020}, and that these transducers can vastly improve the acousto-optic efficiency of nanoscale optomechanical resonators~\cite{Jiang2020}. For many applications, efficient non-resonant acousto-optic transduction is desired.

Here we demonstrate a collinear acousto-optic mode converter using high-confinement, wavelength-scale waveguides in suspended, X-cut films of LN. After reviewing the physics of these modulators including methods for calculating the optomechanical coupling coefficient $g$ (Section~\ref{sec:OM}), we discuss how the optical and mechanical modes can be addressed with AO multiplexers (Section~\ref{sec:multiplexer}). We use these multiplexers to realize a frequency-shifting, four-port optical switch near $1550~\trm{nm}$ and $440~\trm{MHz}$. We describe the behavior of this device in Section~\ref{sec:AO}. The efficiency of the modulator is characterized in Section~\ref{sec:characterization} and used to back out an interaction strength $g/\sqrt{\hbar\Omega}$ of $0.38~\trm{mm}^{-1}\mu\trm{W}^{-1/2}$ which quantifies the required interaction length and mechanical drive power. Owing to this large $g$, this modulator exhibits a record-low power consumption for its length as seen in Table~\ref{tab:FoM}. These modulators are inherently non-reciprocal as demonstrated in Section~\ref{sec:characterization} where we discuss prospects for using them to make non-magnetic isolators. The results reported here mark a significant advance in guided acousto-optics that could enable a new class of low-power, integrated components.

\section{Optomechanics in a waveguide}
\label{sec:OM}

The interactions between light and sound have been studied for a long time~\cite{Brillouin1922}. Here we review these interactions for the modes of a waveguide~\cite{Wolff2015, Sipe2016, VanLaer2016,Eggleton2019}, specifically in the case where sound scatters light between two optical modes. In the context of Brillouin scattering, this is often called \emph{inter-modal}~\cite{Kittlaus2017} or \emph{inter-polarization}~\cite{Kang2011} scattering. In contrast to stimulated Brillouin scattering, here we study how light moves in the presence of a strong mechanical drive, where the light does not affect the dynamics of the sound.

\begin{figure}[h!]
	\includegraphics[width=\linewidth]{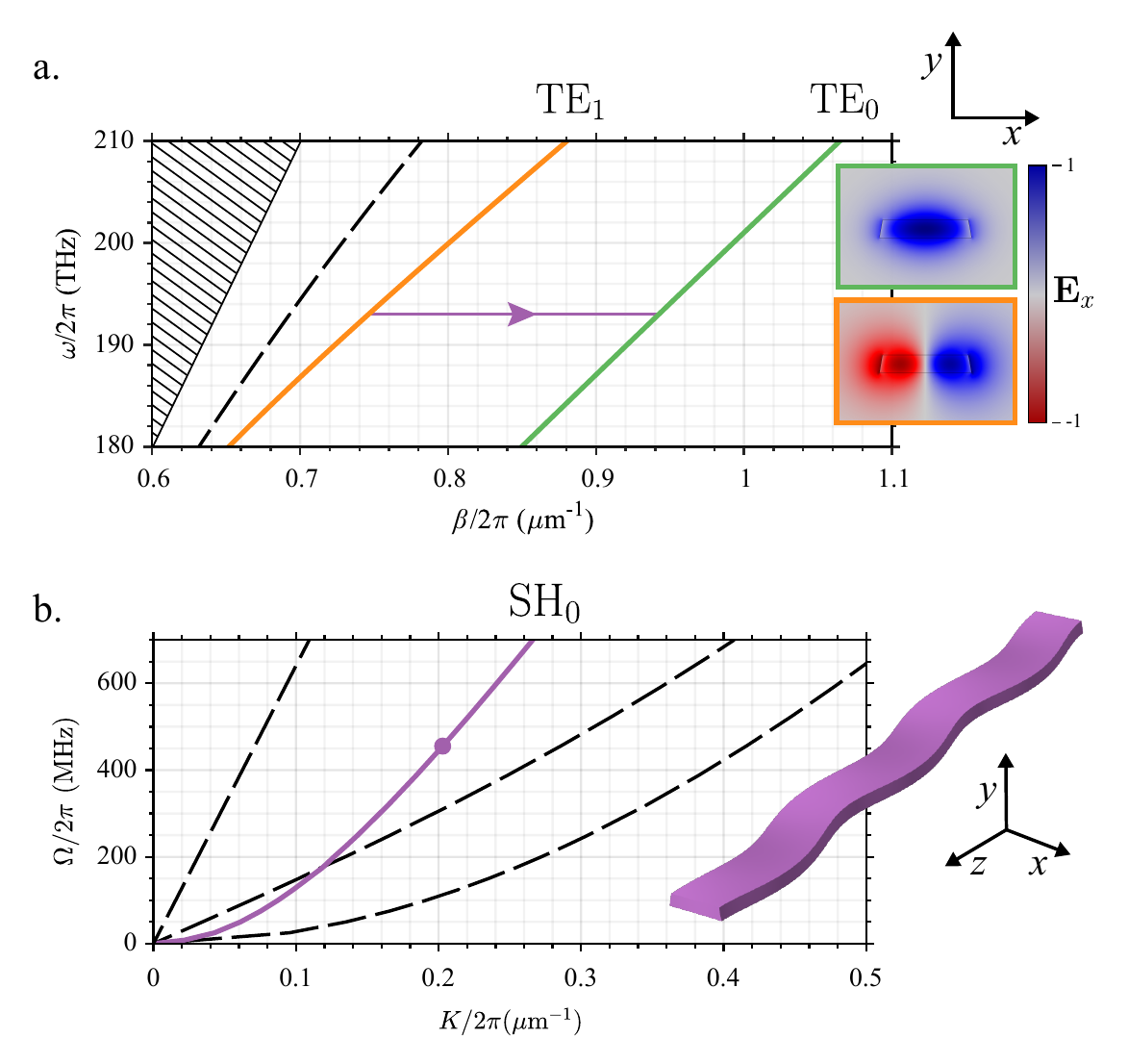}
	\caption{\textbf{Band Structure} \textbf{a.} The LN waveguide investigated supports a TE\(_0\) (green) and TE\(_1\) (orange) mode.  The \(x\)-component of the electric field is plotted showing the TE\(_1\) node on the reflection-symmetry plane. At 193~THz, their wavevectors differ by \(K = 2\pi\times 0.2~\um^{-1}\) shown with the purple arrow. \textbf{b.} The fundamental horizontal shear SH\(_0\) mode with this \(K\) is plotted alongside the mechanical bands (SH\(_0\) in purple). The mode profiles plotted can be used to compute \(g/\sqrt{\hbar\Omega} = 1~\trm{mm}^{-1}\mu\trm{W}^{-1/2}\).}
	\label{fig:modes}
\end{figure}

We direct our attention to Figure~\ref{fig:modes} which shows the optical and mechanical modes of a LN waveguide. The waveguide supports a \TE{0} mode with electric field \(a_0 \E_0 \exp\prens{i\beta_0 z - i \omega_0 t} + \trm{c.c.}\) and a \TE{1} mode with field \(a_1 \E_1 \exp\prens{i\beta_1 z - i \omega_1 t} + \trm{c.c.}\). It also supports the fundamental horizontal shear (SH\(_0\)) mechanical mode with displacement \(b\,\u \exp\prens{iKz - i\Omega t} + \trm{c.c.}\) which scatters light between the \TE{0} and \TE{1} modes. The mode profiles \(\E_i\) and \(\u\) are complex vector fields on the \(xy\)-plane. They are normalized such that \(\left|a_i\right|^2\) and \(\left|b\right|^2\) are the photon and phonon number flux with units of Hz. 

When the scattering process is phasematched, \emph{e.g.}, \(\omega_1 + \Omega = \omega_0\) and \(\beta_1 + K = \beta_0\) as shown in Figure~\ref{fig:modes}a, the coefficients \(\mathbf{a}(t,z) = \prens{a_0(t,z), a_1(t,z)}^\top\) obey
\begin{equation}
	\prens{\mathbf{v}^{-1}\partial_t + \partial_z} \mathbf{a} = -i g b \sigma_x \mathbf{a}
	\label{eq:dynamics}
\end{equation}
as shown in Appendix~\ref{app:waveguideOM}. Here \(\mathbf{v}\) is a diagonal matrix containing the optical group velocities, \(g\) is the optomechanical coupling coefficient, and \(\sigma_x\) is the Pauli-X matrix. In Equation~\ref{eq:dynamics}, we assume $b$ is real.

In the absence of coupling \(g\rightarrow 0\), light in the two modes propagate independently according to a telegrapher equation. When we turn on the coupling, light oscillates between the modes
\begin{align}
	\mathbf{a} &= \exp\prens{-igbz \sigma_x}\mathbf{a}_0 \nonumber\\
		&= \begin{pmatrix}
			 \cos\zeta &  -i\sin\zeta \\
			-i\sin\zeta &  \cos\zeta
		\end{pmatrix}
		\mathbf{a}_0
	\label{eq:ssSoln}
\end{align}
where \(\zeta \equiv gbz\) and \(\mathbf{a}_0\) is vector of the initial coefficients at \(z=0\). These steady-state solutions assume \(b\) is uniform along the waveguide, but we consider a more general case with loss and detuning in Appendix~\ref{app:generalizedDynamics}. Like a bulk acousto-optic modulator, a waveguide-based modulator is a frequency-shifting switch. When \(\zeta = \pi/2\), light initially in mode 1 is converted to mode 2 before being converted back to mode 1 at \(\zeta = \pi\). 

Confining light and sound to a wavelength-scale waveguide enhances the interaction strength \(g\) enabling smaller, more efficient devices. This can be seen in the expression for the coupling coefficient from mode 1 into mode 0
\begin{equation}
  g_{01} = -\frac{\omega_0}{2} \frac{\int\trm{d}A\; \E_{0}^*\delta_\u\perm\cdot \u \E_1}{\mathcal{P}_0 \sqrt{\mathcal{P}_\trm{m}/\hbar\Omega}}.
	\label{eq:g}
\end{equation}
Here $\mathcal{P}_i$ is the optical power in mode $i$ when $|a_i|^2 = 1$, $\mathcal{P}_\trm{m}$ is the mechanical power when $|b|^2 = 1$, and \(\delta_\u\perm\cdot\u\) encodes the permittivity shift from the deformation \(\u\) as described in Appendix~\ref{app:waveguideOM}. First we note how this expression relates to \(g\) in Equation~\ref{eq:dynamics}. By choosing a flux-normalized basis where \(\mathcal{P}_i = \hbar\omega_i\), the coupling takes a Hermitian form \(g_{ij} = g_{ji}^*\). This can be made real and symmetric by choice of phase of the mode profiles, giving us the \(g\) used in equation~(\ref{eq:dynamics})\footnote{If we normalize such that \(|b|^2\) is power instead of flux, the factors of \(\sqrt{\hbar\Omega}\) disappears from \(g\) and throughout the text.}.

Next we consider how \(g\) scales with the area of a waveguide. In the limit of high-confinement, small changes to a waveguide's geometry change its dispersion and the shape of its modes. For fixed $\omega$, $g$ has a complicated dependence on waveguide geometry. When the waves are weakly confined, the numerator in Equation~\ref{eq:g}, $\mathcal{P}_0$, and $\mathcal{P}_\trm{m}$ are approximately proportional to the area of the waveguide $A$. In this regime, the factors of $A$ from the numerator and $P_0$ cancel, leaving only $A^{-1/2}$ from $\mathcal{P}_\trm{m}$. Intuitively, it takes more power \(\mathcal{P}_\trm{m}\) to deform a larger waveguide by \(\u\) which comes at the expense of \(g\) and, ultimately, a device's efficiency. Other three-wave processes like electro-optic and \( \chi^{\prens{2}} \) interactions scale similarly. This motivates the pursuit of high-confinement waveguides for nonlinear and parametric processes like acousto-optics,  underlying recent activity in thin-film LN.

We can use Equation~\ref{eq:g} to calculate the coupling for the rectangular waveguide studied here. Our waveguide is patterned into X-cut LN and suspended by etching the silicon substrate. It is \(1.25~\um\) wide and 250~nm thick. We define the waveguide in a hydrogen silsesquioxane mask and transfer it to the lithium niobate film with an argon ion mill. This produces a \(10^\circ\) sidewall-angle that is included in the simulations. The mode profiles plotted in Figure~\ref{fig:modes} are used to compute the coupling coefficient 
\begin{equation}
	\frac{g}{\sqrt{\hbar\Omega}} = 1.0~\frac{1}{\trm{mm}\sqrt{\mu\trm{W}}}.
\end{equation}
With approximately 40 microwatts of power in the SH\(_0\) mode, the optical wave is completely transfered from the \TE{0} to the \TE{1} mode (or vice versa) after traveling just \(250~\um\) in the waveguide.

\begin{table}[]
    \centering
    \begin{tabular}{|c|c|c|}
        \hline
         Mode & $n_\trm{eff}$ & $n_\trm{g}$ \\
         \hline
         \hline
         \TE{0} & 1.464 & 2.147 \\
         \hline
         \TE{1} & 1.159 & 2.288 \\
         \hline
    \end{tabular}
    \hspace{1em}
    \begin{tabular}{|c|c|c|}
        \hline
         Mode & $2\pi/K$ & $v$ \\
         \hline
         \hline
         SH$_0$ & $5.081~\um$ & $3623~\trm{m/s}$ \\
         \hline
    \end{tabular}
    \caption{The simulated effective and group indices for the optics at 1550~nm, and the wavelength and group velocity for the mechanics that phase-matches these optical modes.}
    \label{tab:modes}
\end{table}

\section{Addressing the optical and acoustic modes of a waveguide}
\label{sec:multiplexer}

\begin{figure}
	\includegraphics[width=\linewidth]{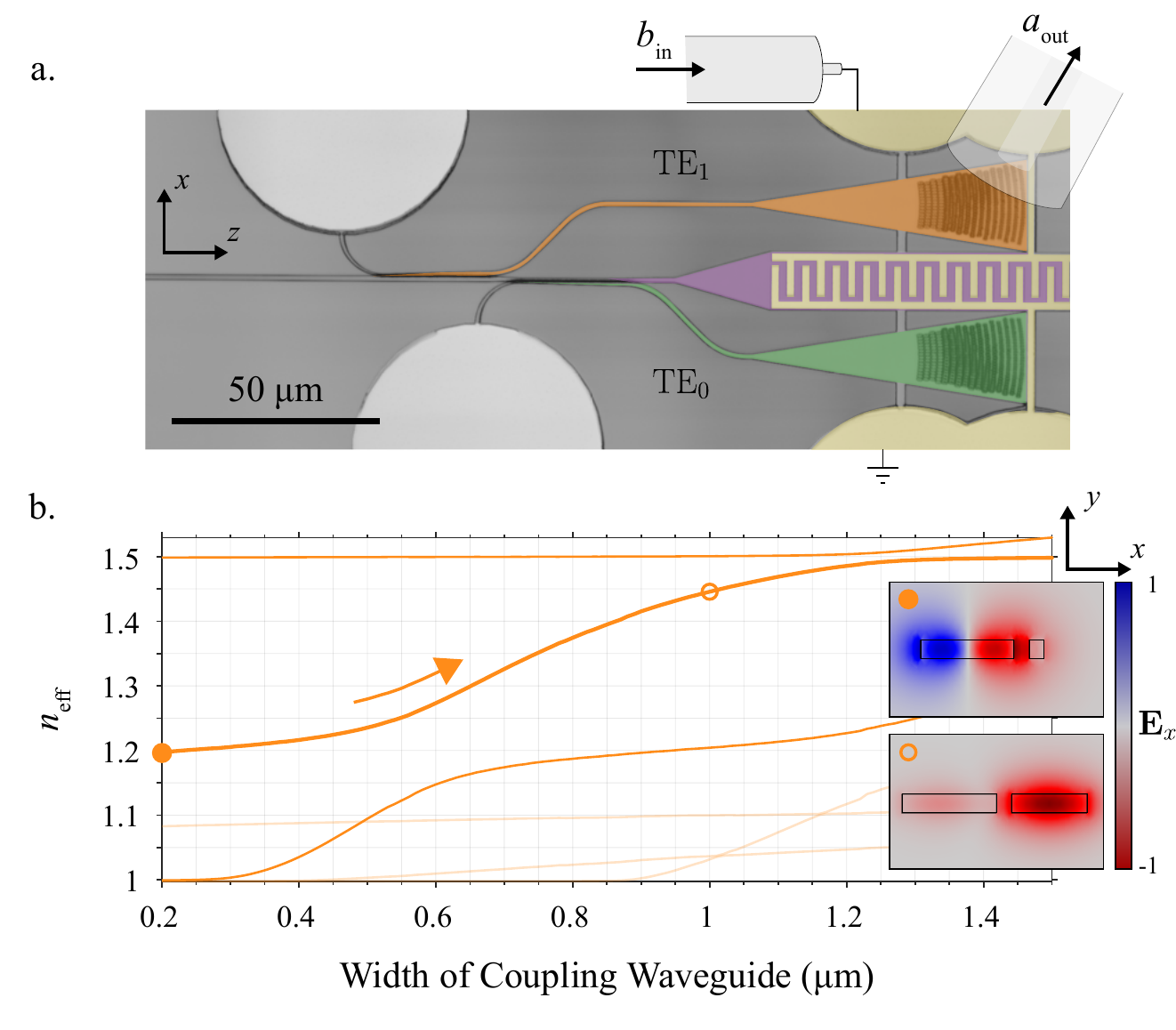}
	\caption{\textbf{AO multiplexer}. \textbf{a.} Light can be injected into the \TE{0} and \TE{1} modes of the waveguide through the green and orange optical ports, respectively. The SH\(_0\) mechanical mode of the waveguide is excited by the purple piezoelectric transducer. \textbf{b.} The optical mode injectors (\TE{1} shown) adiabatically transfer the mode from the waveguide into the \TE{0} mode of the coupler by tapering the width of the coupling waveguide.}
	\label{fig:multiplexer}
\end{figure}

In order to use these interactions to build devices, we need to efficiently address each of the optical and mechanical modes in our waveguide. To this end we engineer the acousto-optic multiplexer~\cite{Dostart2020} shown in Figure~\ref{fig:multiplexer}. This device separates the \TE{0} optical, \TE{1} optical, and  SH\(_0\) acoustic waves into three ports and couples them off-chip.

Light and sound behave very differently at the boundary of a waveguide. Mechanical energy is only transferred between two bodies if they touch. On the other hand, light can tunnel across a gap between adjacent dielectrics. We can use this fundamental difference when designing a multiplexer, evanescently transferring light between adjacent waveguides without disturbing the mechanics.

The optical mode multiplexer comprises two adiabatic tapers. The first on the left of Figure~\ref{fig:multiplexer}a and closer to the AO waveguide couples to the \TE{1} mode and the second to the \TE{0} mode. This design is similar to the cascaded mode-injector developed by Chang \emph{et al.} to use multi-moded waveguides for compact, low-drive power phase shifters~\cite{Chang2017}. In our coupler, the 1.25~\(\um\)-wide AO waveguide supports a \TE{0} and \TE{1} mode. The \TE{1} mode is close to cutoff and has long, evanescent tails. Starting from the left, a 200~nm-wide coupling waveguide is brought in to a distance 200~nm from the AO waveguide. As the width is increased, the \TE{0} mode of the coupling waveguide hybridizes with the \TE{1} mode of the AO waveguide, leading to the anti-crossing in Figure~\ref{fig:multiplexer}b near 600~nm. The coupling waveguide is tapered up to \(1~\um\) in width over \(25~\um\) and the AO waveguide's \TE{1} mode is adiabatically transferred into the coupler's \TE{0}. The coupler waveguide is then bent away from the AO waveguide and sent the top grating coupler in figure~\ref{fig:multiplexer}a. After coupling out \TE{1}, the AO waveguide is narrowed to 575~nm such that \TE{0} exhibits long, evanescent tails and the process is repeated for the \TE{0} mode, this time tapering the coupler from 400~nm to \(1~\um\).

\begin{figure*}[t]
	\includegraphics[width=\textwidth]{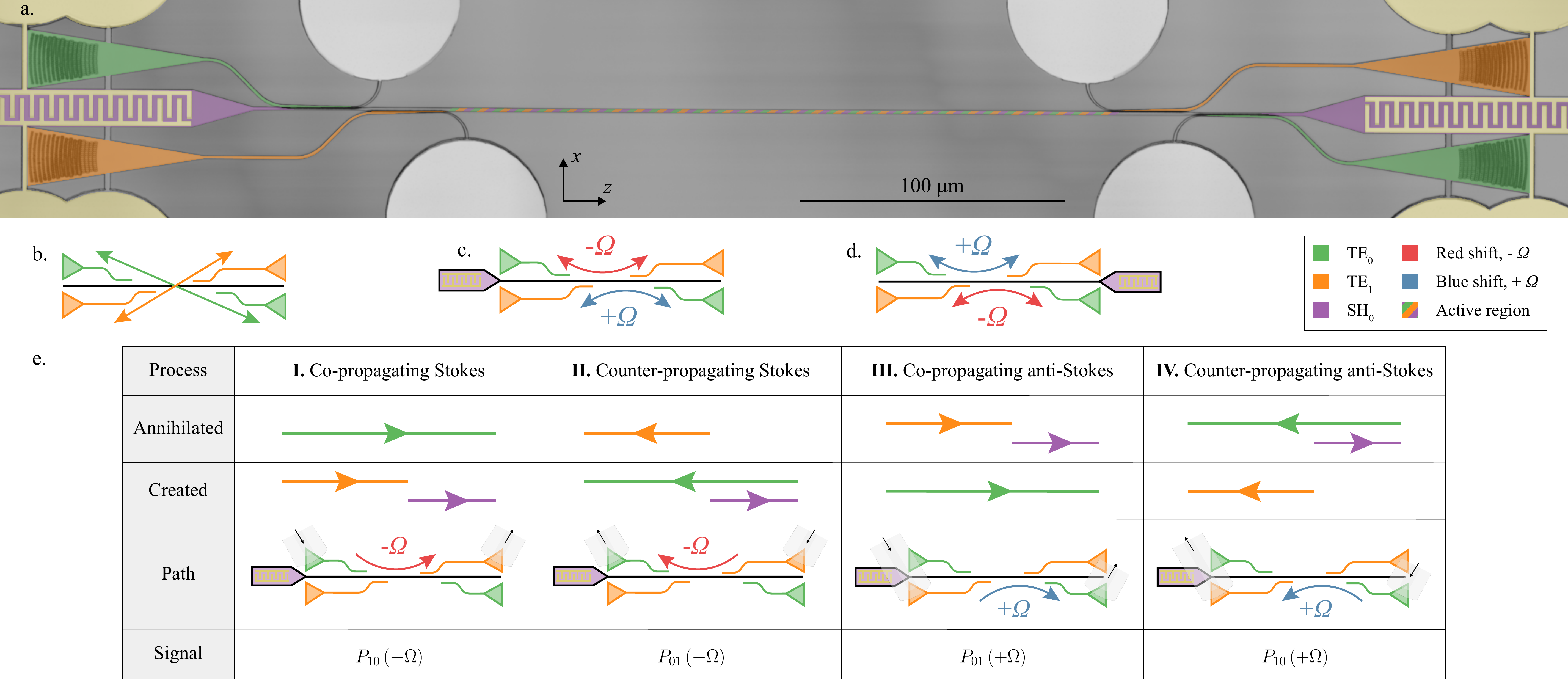}
	\caption{\textbf{Integrated AOM}~\textbf{a.} The full modulator comprises two multiplexers described in Section~\ref{sec:multiplexer} and a waveguide where the interactions happen, labeled ``active region.'' It constitutes a four-optical-port, frequency-shifting switch. \textbf{b.} When no phonons are in the waveguide, light propagates along cross-shaped paths. \textbf{c.} When the left transducer is driven, photons that emit/absorb a phonon travel along the top/bottom bar-shaped path. \textbf{d.} If we change the direction of the phonons, the emission/absorption paths are switched. \textbf{e.} We tabulate the four AO processes that govern the device when driven from the left (\textbf{c.}), including the resulting heterodyne signal as described in Section~\ref{sec:characterization}.}
	\label{fig:device}
\end{figure*}

We measure the optical transmission through the device in Figure~\ref{fig:device}. Our best device exhibited 10~dB isolation, \emph{i.e.}, unintended scatter from the \TE{0} (or \TE{1}) input port to the \TE{1} (\TE{0}) output port. This crosstalk limits the isolation of our AOM. It can be further reduced by optimizing the device, or by actively compensating for crosstalk with an electically tunable feed network~\cite{Miller2015}.

The silicon substrate is etched away, releasing the LN waveguides such that they support optical and mechanical waves. Releasing the device causes the AO and coupling waveguides to deviate from the plane of the chip by different amounts, separating and decoupling the waveguides. To prevent this, we add 150~nm-wide tethers at the ends of each coupling waveguide. These tethers can scatter mechanical waves in the AO waveguide, counter to the design strategy.

Mechanical waves in the waveguide are coupled out with a piezoelectric transducer after the optical couplers. We adapt the transducer design presented in Ref.~\cite{Dahmani2020} for the frequency range of interest. The rescaled design excites the fundamental SH mode with \(K = 2\pi / 5.08~\um^{-1}\) that can phasematch the \TE{0} and \TE{1} optical modes at 1550~nm.

We measure the S-matrix of this two-port microwave system on a calibrated probe station and extract the mechanical propagation loss \(\gamma = -11.7~\trm{dB/mm}\) and the transmission \(|\tbmu|^2 = -21.9~\trm{dB}\). The peak conductance of these transducers is 2.3~mS and, as a result, 4.3~dB of the insertion loss comes from impedance mismatch. The rest is likely from material damping in the transducer~\cite{Dahmani2020}.

A more detailed characterization of the multiplexer, including its frequency response, is presented in  Appendix~\ref{app:multiplexer}.

\section{A waveguide AO modulator}
\label{sec:AO}

The multiplexers described in the previous section give us access to the optical and mechanical modes of the modulator in Figure~\ref{fig:device}. All four phase-matched processes ---~co- and counter-propagating, Stokes and anti-Stokes~--- are at play in this four-port, frequency-shifting switch.

\begin{figure}[h]
	\includegraphics[width=\linewidth]{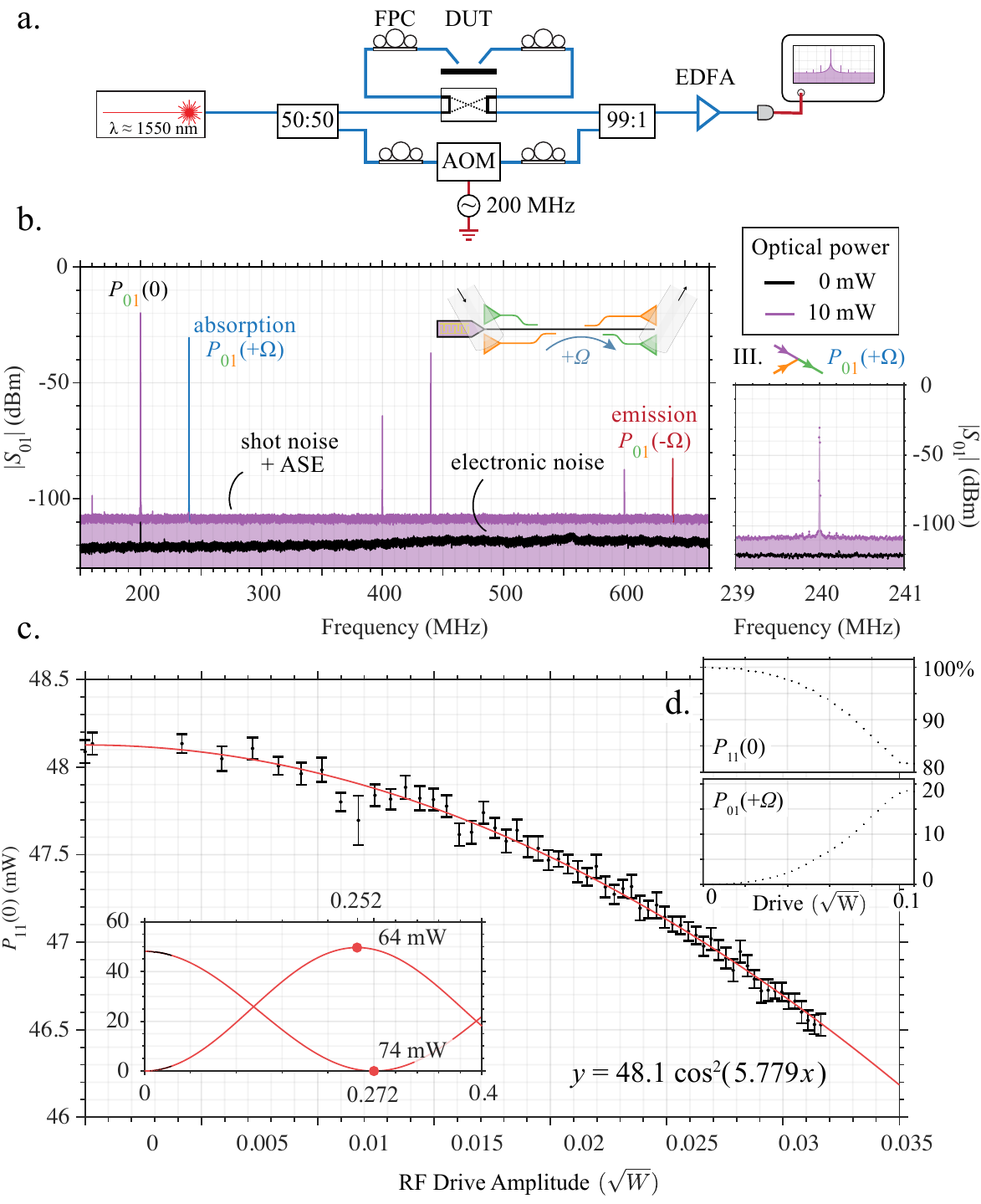}
	\caption{\textbf{Heterodyne measurements}~\textbf{a.} Schematic of the optical heterodyne receiver. FPC: fiber polarization controller. DUT: device under test. EDFA: erbium-doped fiber amplifier. \textbf{b.} An example photocurrent spectrum of the co-propagating anti-Stokes process for a 440~MHz drive. The three tones of interest are the unscattered pump, $P_{01}(0)$ at 200~MHz; the absorption signal, $P_{01}(+\Omega)$ at 240~MHz; and the emission signal, $P_{01}(-\Omega)$ at 640~MHz, which is suppressed by over 50~dB. ASE : amplified spontaneous emission. \textbf{c.} Fitting a sinusoid to the pump depletion $P_{11}(0)$, we extrapolate the full-conversion drive power $P_{\pi/2}$ (inset bottom left) which is used to determine $g$. The absorption signal $P_{01}(+\Omega)$ gives us a similar result. \textbf{d.} At higher drive powers, the signal deviates from a sinusoid (Appendix~\ref{sec:power-handling}). We observe up to 18\% conversion, where the signals have been normalized by the undepleted pump ($P_{11}(0)$ with the drive off). }
	\label{fig:heterodyne}
\end{figure}

First we consider what happens without the mechanical drives. We send light at \(\omega\) into the bottom-left grating (orange) which gets injected into the \TE{1} mode of the waveguide. With no phonons in the waveguide, the light passes through the device and gets removed by the \TE{1} injector, leaving the chip from the top-right grating (orange). The optical paths through an undriven device are shown in Figure~\ref{fig:device}b.

Now consider the co-propagating anti-Stokes process (Figure~\ref{fig:device}e.~III). Again sending light into the bottom-left, we drive the left transducer with an RF tone at \(\Omega\). This sends phonons down the waveguide to the right. Photons in the \TE{1} mode of the waveguide absorb co-propagating phonons and scatter into the \TE{0} mode. After absorption, their frequency increases to \(\omega + \Omega\) and their wavevector increases from \(\beta_1(\omega)\) to \(\beta_0(\omega+\Omega) = \beta_1(\omega) + K\)  (also shown in Figure~\ref{fig:modes}). The \TE{0} injector removes these up-shifted photons from the waveguide and they are scattered off-chip by the grating in the bottom-right (green).

If we instead send light into \TE{0} from the top-left (green), the co-propagating phonons stimulate emission --- instead of absorption --- and the incident light scatters into the \TE{1} mode at \(\omega - \Omega\). The down-shifted light leaves the chip through the top-right grating (orange). This is the co-propagating Stokes process diagrammed in Figure~\ref{fig:device}e.~I.

In addition to the two co-propagating processes described above, there are counter-propagating processes which we probe by sending the optical field from right-to-left (Figure~\ref{fig:device}e.~II, IV). The four processes are summarized in Figure~\ref{fig:device}c. The co-propagating and counter-propagating Stokes processes --- e. I and e. II, respectively --- form the top, red-shifted path. The co- and counter-propagating anti-Stokes processes --- III and IV --- form the bottom, blue-shifted path. 
When $L_\trm{eff}\gg \Omega^{-1}(v_1^{-1} + v_0^{-1})^{-1}$, the co- and counter-propagating processes are not simultaneously phase-matched.

Driving the mechanics from the right, \emph{i.e.}, flipping the direction of the phonon, also switches absorption and emission. This gives us Figure~\ref{fig:device}d which, because of the symmetry of our device, is equivalent to \ref{fig:device}c under a $180^\circ$ rotation.

We can think of the device as a frequency-shifting optical switch. No matter which direction the phonons are coming from, the mechanical drive switches the device from the ``cross'' state (b) to the ``bar'' state (c and d). The direction of the mechanical wave determines which path red-shifts and which path blue-shifts the light.

\section{Characterizing the modulation}
\label{sec:characterization}

In addition to the mechanical efficiency and attenuation reported in Section~\ref{sec:multiplexer}, $g$ is a key figure that determines the modulator's efficiency. We determine \(g\) by measuring the scattered power and pump depletion using the heterodyne setup in Figure~\ref{fig:heterodyne}a. Light in the telecom C band is generated with a Santec TSL-550 laser. It is split with half the light sent to the device and the other half up-shifted by \(\Delta \equiv 2\pi \times 200~\trm{MHz}\) using an AOM. The paths are recombined with a 99:1 splitter and sent to an Optilab PD-40-M detector. The photocurrent spectrum \(S_{II}[\omega]\) is measured with a Rhode \& Schwartz FSW.

A photocurrent spectrum of the co-propagating anti-Stokes process (Figure~\ref{fig:device}~e.~III) is shown in Figure~\ref{fig:heterodyne}b for a drive frequency $\Omega = 2\pi\times 440~\trm{MHz}$. There are three important tones in the spectrum : one at $\Delta$ and the other two at $\Omega \pm \Delta$. If the light travels through the device without being scattered by the acoustic wave, its frequency stays the same, leading to the RF tone at 200~MHz. If the light absorbs/emits a phonon from the acoustic field, its frequency is shifted by $\pm 440~\trm{MHz}$, generating the tone at 240/640~MHz. We denote the power in these tones $P_{ij}(\omega^\prime)$ where $ij$ specifies the optical path \TE{j}$\rightarrow$\TE{i} and $\omega + \omega^\prime$ is the frequency of the light generating the tone. The power,
\begin{equation}
  P_{01}(+\Omega) = Z_0\int_{-B/2}^{B/2}\trm{d}\omega\; S_{II}\left[\Omega - \Delta + \omega\right]
\end{equation}
for our example process, is proportional to the optical power emitted from the device. Here \(Z_0\) is 50~Ohms and \(B\) is the integration bandwidth. Each process in Figure~\ref{fig:device}e is labeled with the resulting $P_{ij}(\omega^\prime)$.

By fitting the scattered power $P_{10}(+\Omega)$ and pump depletion $P_{11}(0)$, we determine \(g\) in way that is insensitive to the calibration of the loss in the optical signal chain (Appendix~\ref{app:inferring_g}). For a phase-matched process, these fits (Figure~\ref{fig:heterodyne}c) give us \(g\tbmu L_\trm{eff}/\sqrt{\hbar\Omega}\) where
\begin{equation}
    L_\trm{eff} = 2\gamma^{-1}\left\lbrack 1 - \exp\prens{-\gamma L / 2} \right\rbrack
\end{equation}
as described in Appendix~\ref{app:generalizedDynamics}. In Figure~\ref{fig:heterodyne}c, we extrapolate \(P_{\pi/2}\), the power it takes to fully swap \TE{1}$\leftrightarrow$\TE{0}, which is related to $g$ as
\begin{equation}
    g\tbmu L_\trm{eff}\sqrt{\frac{P_{\pi/2}}{\hbar\Omega}} = \frac{\pi}{2}.
\end{equation}
After removing the cable loss (0.6~dB), we find 
\begin{equation}
	\frac{g}{\sqrt{\hbar\Omega}} = 0.38~\frac{1}{\trm{mm}\sqrt{\mu\trm{W}}},
\end{equation}
roughly a third the simulated value. This discrepancy is similar to what we have observed for waveguides in LN films on sapphire~\cite{Sarabalis2020} and resonant optomechanical systems in suspended LN films~\cite{Jiang2019}. 

To better understand the different device architectures, we extend the table presented by Smith~\emph{et al.}~\cite{Smith1990} to include $\tbmu$, $g$, and recent work with high-confinement waveguides (Table~\ref{tab:FoM}). If the conversion efficiency $P_{\pi/2}$ is not directly measured, we extrapolate $P_{\pi/2}$ from low-power measurements of the efficiency. The transducer's transmission $\tbmu$ is not typically reported. Here we de-embed $\tbmu$ from the two-port S-parameter (Appendix~\ref{sec:mechanics}). Without careful analysis, $\tbmu$ is prone to over-estimate. For example, Rayleigh and Bleustein-Gulyaev waves are nearly degenerate for X-cut LN devices~\cite{Heffner1988,Frangen1989,Hinkov1991,Hinkov1994} and, if not properly handled, degrade $\tbmu$ without reducing $|S_{21}|$~\cite{Duchet1995}. Where we are unable to infer $\tbmu$, we assume $\left|\tbmu\right|^2 + \left|S_{11}\right|^2 = 1$, which strictly over-estimates $|\tbmu|$. The optomechanical coupling coefficient is inferred from $\tbmu$ and $P_{\pi/2}$ taking into account mechanical loss $\gamma$ where reported. The figure-of-merit $(L/\lambda)^2 P_{\pi/2}$ is adapted from Smith~\emph{et al.}~\cite{Smith1990}. It is proportional to $\left|g\tbmu\right|^{-2}$. A low-power, compact modulator requires a low figure-of-merit, \emph{i.e.}, both an efficient transducer and strong photon-phonon interactions.

\begin{table*}[]
    \centering
    \begin{tabular}{|c|c|c|c|c|c|c|c|c|c|c|}
        \hline
        Work & Year & $\lambda$ & $\Omega/2\pi$ & $L$ & $P_{\pi/2}$ & $\gamma$ & $\left|\tbmu\right|^2$ & $g / \sqrt{\hbar\Omega}$ & $(L / \lambda)^2 P_{\pi/2}$ & \(\eta_\trm{max}\) \\
          &  & (nm) & (MHz) & (mm) & (mW) & (dB/mm) & (dB) & (mm$^{-1}$ W$^{-1/2}$) & (MW) & (\%) \\ 
        \hline\hline
        Harris~\cite{Harris1970} & 1970     & 632.8 & 54       & 35     & $3.64 \times 10^3$    & ---   & -7.5  & 0.057     & $1.1\times 10^4$    & 95         \\
        \hline                                                           
        Ohmachi~\cite{Ohmachi1977} & 1977   & 1150  & 245.5    & 4.5    & 550                   & ---   & -20   & 4.7       & 8.42                & 70           \\
        \hline                                                           
        Binh~\cite{Binh1980}      & 1980    & 632.8 & 550      & 9      & 225                   & ---   & -25   & 6.54      & 45.5                & 99          \\
        \hline                                                           
        Heffner~\cite{Heffner1988}   & 1988 & 1523  & 175      & 25     & 500                   & ---   & -7.0  & 0.20       & 135                 & 97          \\
        \hline                                                           
        Hinkov~\cite{Hinkov1988} & 1988     & 633   & 191.62   & 17     & 400                   & -0.1  & -25   & 2.6      & 289                 & 90           \\
        \hline                                                           
        Frangen~\cite{Frangen1989}   & 1989 & 1520  & 178      & 9      & 90                    & -0.05 & -10   & 1.9      & 3.16                & 99          \\
        \hline                                                           
        Hinkov~\cite{Hinkov1991}    & 1991  & 800   & 355.5    & 20     & 19.8                  & -0.04 & -3    & 0.825     & 12.4                & 93          \\
        \hline                               
        Hinkov~\cite{Hinkov1994}    & 1994  & 800   & 365      & 25     & 0.5                   & -0.04\tablefootnote[1] & -3\tablefootnote[1] & 4.2       & 0.488               & 100             \\
        \hline                               
        Duchet~\cite{Duchet1995}    & 1995  & 1556  & 170      & 30     & 6                     & ---   & -3    & 0.96      & 2.23                & 100             \\
        \hline                               
        \hline                               
        Liu~\cite{Liu2019}       & 2019     & 1510  & 16,400    & 0.5   & $4.2 \times 10^5$     & ---   & -15   & 0.041     & 46.4                & $2.5\times 10^{-4}$ \\
        \hline
        Kittlaus~\cite{Kittlaus2020}\tablefootnote[2]  & 2020 & 1600 & 3,110     & 0.240 & $4.69 \times 10^3$    & ---  & -12   & 5.7       & 0.105               & 1          \\
        \hline
         ''   &  ''  &  1525.4 & ''  & 0.960 & $1.48 \times 10^3$    & ---   & ''  & 4.5       & 0.587               & 13.5      \\
        \hline
        This work & 2020 & 1550 & 440      & 0.25  & 60                    & -11.7  & -21.9 & 377       & $1.6\times 10^{-3}$ & 18          \\
        \hline
    \end{tabular}
    \footnotetext[1]{Value copied from Ref.~\cite{Hinkov1991}.}
    \footnotetext[2]{The two rows are for the straight (top) and serpentine (bottom) modulators reported. These devices are side-coupled, not collinear.}
    \caption{\textbf{Literature Review} We extend the table compiled by Smith~\emph{et al.}~\cite{Smith1990} to include estimates for $\tbmu$ and $g$, and recent work on high-confinement waveguide devices. $\eta_\trm{max}$ is the maximum conversion efficiency demonstrated. The values for $g$ are inferred from $P_{\pi/2}$. We account for mechanical propagation loss where reported. High-confinement modulators dramatically improve the figure-of-merit $(L/\lambda)^2 P_{\pi/2}$.}
    \label{tab:FoM}
\end{table*}

\begin{figure}[h]
	\includegraphics[width=\linewidth]{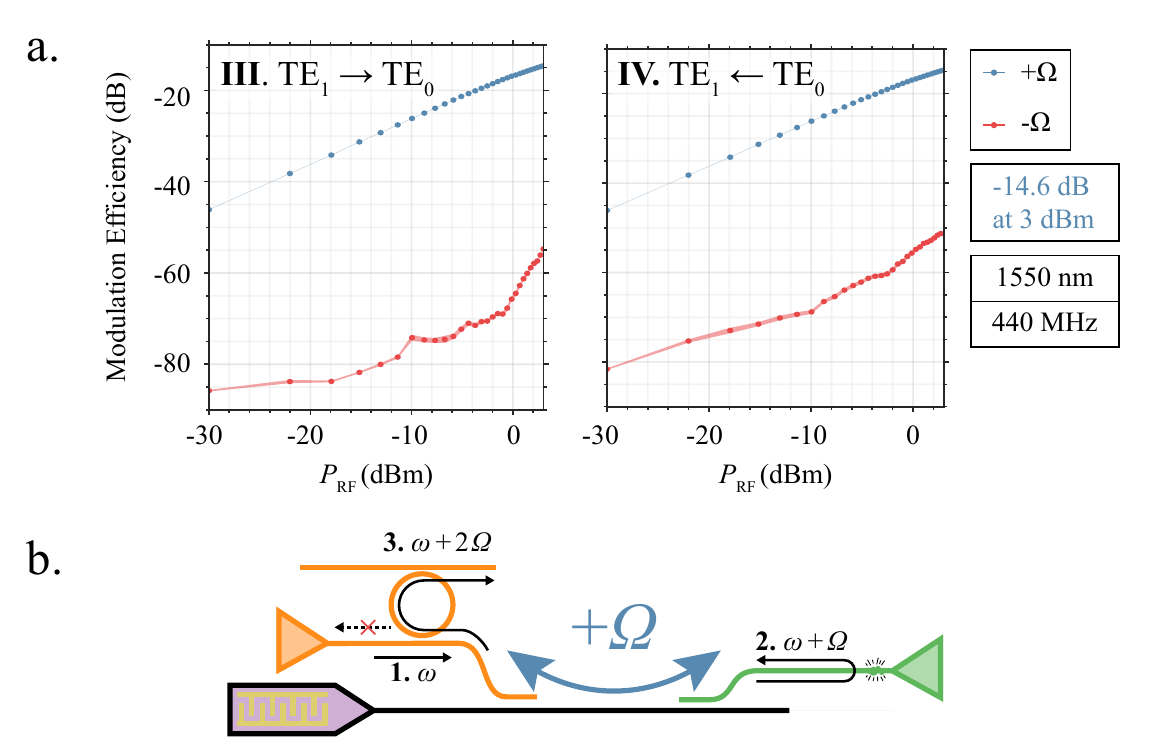}
	\caption{\textbf{Nonreciprocity} \textbf{a.} Light traveling along the bottom path of Figure~\ref{fig:device}c incurs a $+\Omega$ blue-shift independent of the direction it travels. This is a signature of nonreciprocity. Here \emph{modulation efficiency} is the scattered power over the undepleted pump, \emph{e.g.}, $P_{01}(+\Omega)$ over $P_{11}(0)$ for III. \textbf{b.} The resulting round-trip frequency-shift can be used to build a frequency-shifting isolator. A ring filter can be added before the AOM to drop reflections.}
	\label{fig:nonreciprocity}
\end{figure}

Acousto-optic modulators are inherently nonreciprocal and can be used to make isolators and circulators to stabilize lasers~\cite{Smith1973} and help manage reflections in large photonic circuits. Nonreciprocal components usually employ the magneto-optical effect, motivating the pursuit of thin-film YIG functional layers in silicon photonics~\cite{Tien2011,Ghosh2012}. Alternatively, parametric drives like acousto-optics give us a non-magnetic way to build nonreciprocal components ~\cite{Smith1973,Heeks1982,OMeara1988,Li2014,Peano2015,Sohn2018, Sarabalis2018,Sohn2019,Kittlaus2020, Williamson2020}.

In these devices, nonreciprocity takes a slightly different form than in a standard isolator. Consider the bottom path in Figure~\ref{fig:device}c. Light traveling between the \TE{1} and \TE{0} port absorbs a phonon independent of the direction it travels. This can be seen directly in the path-independent blue-shift measured in Figure~\ref{fig:nonreciprocity}a. If light takes a round-trip through the device, it absorbs two phonons and returns to a different state.

This round-trip frequency shift is nonreciprocal. It can be used to build the frequency-shifting isolator in Figure~\ref{fig:nonreciprocity}b~\cite{Heeks1982}. Light back-scattered from, \emph{e.g.}, an imperfect component is shifted by $+2\Omega$ and, as a result, can be dropped by a filter to isolate the input port from reflections. When $L_\trm{eff} \gg v_1/2\Omega$, the backwards process is not phase-matched and the reflections are dropped even without a filter~\cite{Kittlaus2020}. 
Finally, this device can be cascaded with another AOM to make a fixed-frequency isolator~\cite{OMeara1988}.

\section{Outlook}

Building off our recent work on waveguide transduction~\cite{Dahmani2020}, we demonstrate a compact acousto-optic modulator in a suspended film of X-cut lithium niobate. The modulator comprises a frequency-shifting, four-port optical switch at $440~\trm{MHz}$ and $1550~\trm{nm}$. By employing high-confinement optical and mechanical modes with strong photon-phonon interactions ($g/\sqrt{\hbar\Omega} = 0.38~\trm{mm}^{-1}\mu\trm{W}^{-1/2}$), it exhibits a record-setting figure-of-merit. This device is optically broadband and, as demonstrated, inherently non-reciprocal. We discuss how it can be integrated with an optical filter to implement a fixed-frequency, non-magnetic isolator.

High-confinement waveguides mark a significant advance in the figure-of-merit of collinear AOMs. The device reported here offers a two order-of-magnitude increase in $g$ and a $1,000\times$ improvement in the figure-of-merit over Ti-indiffused and proton-exchanged waveguides. While high-confinement AOMs have yet to reach full conversion or a record $P_{\pi/2}$, both are within reach. Piezoelectric waveguide transducers similar to the ones used here but centered at $2~\trm{GHz}$~\cite{Dahmani2020} are nearly 10~dB more efficient. Recovering that 10~dB --- for example, through design improvements --- would yield full conversion. With a modest length increase (\emph{e.g.}, $L = 1~\trm{mm}$), $P_{\pi/2}$ would drop below a milliwatt. 

Reducing the loss in these high-confinement mechanical waveguides could dramatically lower the necessary drive power. The mechanical loss measured here, $\gamma = -11.7~\trm{dB/mm}$, limits $L_\trm{eff}$ to $742~\um$. At similar frequencies, low-confinement devices exhibit a $\gamma$ as small as $-0.04~\trm{dB/mm}$~\cite{Hinkov1991} and therefore $L_\trm{eff}$ as long as 22~cm. Recently high-confinement mechanical waveguides have been demonstrated with $\gamma = 0.05~\trm{dB/mm}$~\cite{Fu2019}. Increasing $L_\trm{eff}$ from $250~\um$ to 1~cm/10~cm decreases $P_{\pi/2}$ by -32~dB/-52~dB with another $-22$~dB available from improvements to $\tbmu$. Without increasing $g$, $P_{\pi/2}$ could be as low as a nanowatt. 

A long effective length not only improves efficiency, it is necessary for narrow bandwidth optical filtering. A typical commercial AO tunable filter offers nm-scale bandwidth. The 3~cm length of the AO tunable filter in Reference~\cite{Hinkov1991} enabled a bandwidth of 0.32~nm. For high-confinement waveguides to realize compelling filter functions, we need either better mechanical waveguides or to distribute the mechanical transduction along the waveguide (\emph{e.g.}, multiple side-coupled transducers in the architecture in Reference~\cite{Kittlaus2020}). 

Lastly, switching to an unsuspended platform presents opportunities to integrate AO components into larger, more complex circuits and systems which draw from the growing toolbox of piezoelectric, electrooptic, nonlinear, and even quantum components in thin-film LN. One candidate is lithium niobate-on-sapphire which exhibits good acousto-optic and piezoelectric properties~\cite{Sarabalis2020}, and in which efficient waveguide transducers have been recently demonstrated~\cite{Mayor2020}.

The acousto-optic modulator presented here marks an advance of rapidly developing waveguide technology and material platform. Acousto-optic devices like this could soon play a role as compact, low-power frequency-shifters, non-magnetic isolators, tunable filters, and beam deflectors in complex circuits and systems.

\section{Acknowledgements}

This work was supported by a MURI grant from the U. S. Air Force Office of Scientific Research (Grant No. FA9550-17-1-0002), the DARPA Young Faculty Award (YFA), by a fellowship from the David and Lucille Packard foundation, and by the National Science Foundation through ECCS-1808100 and PHY-1820938. The authors wish to thank NTT Research Inc. for their financial and technical support. Part of this work was performed at the Stanford Nano Shared Facilities (SNSF), supported by the National Science Foundation under Grant No. ECCS-1542152, and the Stanford Nanofabrication Facility (SNF).

\section{Author Contributions}

C.J.S. led the project and wrote the manuscript with help from R.V.L. and A.H.S.-N.. R.V.L. and R.N.P. contributed to the measurements; Y.D.D. to the development of the piezoelectric transducers; and W.J. and F.M.M. to the development of the fabrication process. A.H.S.-N. supervised the project.

\appendix

\section{Optomechanics in a waveguide}
\label{app:waveguideOM}

For the ease of the reader, in this Appendix  we derive the coupled mode theory in Equation~\ref{eq:dynamics} including the form for the coupling coefficient Equation~\ref{eq:g} from Maxwell's equations. The coupled mode equations used here are essentially the same as those derived by Yariv in 1973~\cite{Yariv1973} and are a special case of the equations of motion used in the Brillouin scattering literature. In the Brillouin literature, the dynamics of the mechanical field $b$ is on equal-footing with the optical fields $a_i$. Here, as is appropriate for an acousto-optic modulator, we assume a strong mechanical drive such that $b$ is approximately unaffected by the optical fields. $b$ appears as a non-dynamical parameter in the equations of motion for $a_i$.

For Brillouin scattering, Sipe and Steel provide a derivation of the coupled mode theory starting with the Hamiltonians for elasticity, electromagnetism, and their parametric optomechanical coupling~\cite{Sipe2016}. This approach allows them to derive a fully quantum theory. Wolff \emph{et al.} derive the classical coupled mode theory from the second-order differential equations of motion for the electric field and mechanical displacement field~\cite{Wolff2015}. Our approach is similar to a special case of that of Wolff \emph{et al.} but is built off the first-order differential form of Maxwell's equations.

\subsection{Coupled-mode theory from Maxwell's Equations}

Maxwell's equations which generate the motion of the field
\begin{align}
    \nabla\times\mathbf{E}  &= -\partial_t\mu\mathbf{H} \\
    \nabla\times \mathbf{H} &= \partial_t \perm\mathbf{E}
\end{align}
can be expressed compactly
\begin{equation}
    \nabla\times i\sigma_y \Psi = \partial_t \Pi\Psi
    \label{eq:MEcompact}
\end{equation}
by defining the two-component field vector
\begin{equation}
    \Psi = 
        \begin{pmatrix}
            \mathbf{E} \\
            \mathbf{H}
        \end{pmatrix}
\end{equation}
and the matrix
\begin{equation}
    \Pi = 
        \begin{pmatrix}
            \perm   & 0 \\
            0       & \mu 
        \end{pmatrix}
        .
\end{equation}
A waveguide with continuous translation symmetry along $z$ has solutions of the form
\begin{equation}
    \Psi\prens{t,x,y,z} = \psi\prens{x,y} e^{i\beta z - i\omega t}.
    \label{eq:waveguideModes}
\end{equation}
For a given $\omega$, the mode profile $\psi\prens{x,y}$ and wavevector $\beta$ solve the eigenvalue problem
\begin{equation}
    \prens{\nabla_\perp \times i\sigma_y + i\beta \hat{z}\times i\sigma_y + i\omega\Pi} \psi = 0
    \label{eq:waveguideEigenvalueProblem}
\end{equation}
following from Equation~\ref{eq:MEcompact}.

The modes of a waveguide are orthogonal under two inner products as shown in Sections~\ref{sec:power-orthogonality} and \ref{sec:energy-orthogonality}. Consider two solutions with profiles $\psi_i$ and $\psi_j$ and wavevectors $\beta_i \ne \beta_j$. If $\Pi$ is Hermitian, \emph{i.e.}, $\Pi = \Pi^\dagger$, the profiles are power-orthogonal
\begin{align}
    -2\mathcal{P}_{ij} &= \int\trm{d}A\;\psi_i^* \cdot \mathbf{\hat{z}}\times i\sigma_y\psi_j \\
    &= 0
    \label{eq:power-orthogonality}
\end{align}
where $\mathbf{\hat{z}} = \prens{\hat{z},\hat{z}}^\top$. In terms of the fields,
\begin{align}
    \mathcal{P}_{ij} = \frac{1}{2}\hat{z}\cdot\int\trm{d}A\, \mathbf{e}_i^* \times \mathbf{h}_j + \mathbf{e}_j\times\mathbf{h}_i^*
\end{align}
where $\mathbf{e}\prens{x,y}$ and $\mathbf{h}\prens{x,y}$ are the electric and magnetic field profiles, respectively. When $i = j$, this is the z-component of the time-averaged power. Similarly if $\Pi = \Pi^*$ and $[\sigma_z,\Pi] = 0$, the profiles are also energy-orthogonal
\begin{align}
    2\mathcal{E}_{ij} &= \int\trm{d}A\,\psi_i^*\cdot\Pi\psi_j \\ 
        &= 0
        \label{eq:energy-orthogonality}
\end{align}
where
\begin{equation}
    \mathcal{E}_{ij} = \frac{1}{2}\int\trm{d}A\, \mathbf{e}_i^*\perm\mathbf{e}_j + \mathbf{h}_i^*\mu\mathbf{h}_j
\end{equation}
is the energy density along $z$ when $i = j$.

The modes of the waveguide give us a basis in which we can express an arbitrary harmonic solution. Now we show, following from our orthogonality relations, that this basis diagonalizes the dynamics yielding a system of independent telegrapher equations. We are primarily interested in the dynamics of waves in a narrowband about $\omega$. In this case, we can decompose the field in our basis of waveguide modes
\begin{equation}
    \Psi\prens{t,x,y,z} = \sum_i a_i\prens{t,z}\psi_i\prens{x,y}e^{i\beta_i z - i\omega t}
    \label{eq:modeBasis}
\end{equation}
and find the dynamics of the coefficients, or ``envelopes,'' $a_i(t,z)$. The constraint on the bandwidth of $a_i(t,z)$ is the ``slowly varying envelope approximation.''  Sipe and Steel describe how higher order corrections to the field can be included in the dynamics~\cite{Sipe2016}. Equation~\ref{eq:modeBasis} has a sum over discrete bands but can be generalized to include an integral over a continuum of states, like the radiative modes in the air surrounding our LN waveguide.

Each of the modes $\psi_i \exp\prens{i\beta_i z -i\omega t}$ solves Maxwell's equations and so, if we expand Equation~\ref{eq:MEcompact} in this basis, only the derivatives acting on the coefficients $a_i(t,z)$ remain. Acting on each side with $\int\trm{d}A\;\psi_i^*\cdot$, we use our orthogonality relations to find the equation of motion for $a_i$
\begin{equation}
    \prens{-2\mathcal{P}_{ii}\partial_z - 2\mathcal{E}_{ii}\partial_t}a_i\prens{t,z} = 0.
\end{equation}
Defining the group velocity $v_i \equiv \mathcal{P}_{ii}/\mathcal{E}_{ii}$, we can re-express this as
\begin{equation}
    \prens{v_i^{-1}\partial_t + \partial_z} a_i\prens{t,z} = 0.
\end{equation}
Finally defining a vector $\mathbf{a}$ with components $a_i$ and matrix $\mathbf{v}$ with diagonal component $v_i$, we have
\begin{equation}
    \prens{\mathbf{v}^{-1}\partial_t + \partial_z} \mathbf{a}\prens{t,z} = 0.
    \label{eq:telegrapher}
\end{equation}
In these equations, light in each mode propagates at the mode's group velocity. In the next section, we incorporate physics which modulates and couples the modes.

\subsection{Perturbative coupling and inter-modal scattering in optomechanics}

Many of the effects in parametrically driven systems and nonlinear optics can be captured in the coupled mode theory by including a polarization drive field to the RHS of the equations~\cite{Yariv1973}. Optomechanics, electro-optics, and thermo-optics are all examples of parameteric modulation in which the drive field comes from perturbing the material $\Pi \rightarrow \Pi + \delta\Pi$ such that
\begin{equation}
    \prens{v_i^{-1}\partial_t + \partial_z} a_i\prens{t,z} = -\frac{\int\trm{d}A\,\prens{\psi_1e^{i\omega t - i\beta_1 t}}^*\cdot \partial_t\delta\Pi \psi}{2\mathcal{P}_{ii}}.
    \label{eq:dynamicsWithPerturbation}
\end{equation}
For optomechanics, a mechanical field $\mathbf{u}$ perturbs $\perm$ such that
\begin{equation}
    \perm \rightarrow \perm + \delta_\mathbf{u}\perm\cdot\mathbf{u}.
\end{equation}
The perturbation has a radiation pressure term that is delta-distributed on boundaries between dielectrics $\partial R$~\cite{Johnson2002} and a photoelastic term~\cite{Andrushchak2009}
\begin{align}
    \delta_\mathbf{u}\perm\cdot\mathbf{u} = &\prens{\Delta\perm\mathbf{\Pi}_\shortparallel - \perm\Delta\perm^{-1}\perm\mathbf{\Pi}_\perp }\prens{\mathbf{u}\cdot\hat{n}}\delta_{\partial R}  \\ 
     & - \perm \cdot \frac{p \mathbf{S}}{\perm_0} \cdot \perm.
\end{align}
Here $\hat{n}$ is normal to the boundary $\partial R$ pointing from dielectric $1$ into dielectric $2$; $\Delta\perm$ and $\Delta\perm^{-1}$ are $\perm_2 - \perm_1$ and $\perm^{-1}_2 - \perm^{-1}_1$; $\mathbf{\Pi}_\perp$ and $\mathbf{\Pi}_\shortparallel$ project the field perpendicular and parallel to $\hat{n}$, respectively; $p$ is the photoleastic tensor; and $S_{ij} = (\partial_i u_j + \partial_j u_i)/2$ is the strain in the deformed medium.  

With our expression for the perturbation $\delta\Pi$ for optomechanics, we turn our attention to the case treated in the manuscript : coupling between the \TE{0} mode with amplitude $a_0$ and the \TE{1} with amplitude $a_1$. 

Only phase-matched interactions contribute constructively over long interaction times and distances. Consider the mechanical wave
\begin{equation}
    \mathbf{u}\prens{t,x,y,z} = \sqrt{\Phi_\trm{m}^{-1}}\prens{b \mathbf{u}\prens{x,y} e^{iKz-i\Omega t} + \trm{c.c.}}
    \label{eq:mechanicalWave}
\end{equation}
propagating along $+z$. The amplitude $b$ is taken to be real and constant. The mode is flux normalized where $\Phi_\trm{m} = \mathcal{P}_\trm{m}/\hbar\Omega$ such that $\left|b\right|^2$ is the phonon flux with units of Hz. If the mechanical mode phase-matches the \TE{0} and \TE{1} modes, \emph{i.e.},  $\omega_0 = \omega_1 + \Omega$ and $\beta_0 = \beta_1 + K$, Equation~\ref{eq:dynamicsWithPerturbation} for the \TE{0} amplitude becomes 
\begin{equation}
    \prens{v_1^{-1}\partial_t + \partial_z} a_0\prens{t,z} = -i g_{01} b a_1
\end{equation}
with
\begin{equation}
    g_{01} = -\frac{\omega_0}{2}\frac{\int\trm{d}A\,\mathbf{e}_0^*\cdot \delta_\mathbf{u}\perm\cdot\mathbf{u}\cdot\mathbf{e}_1 }{\mathcal{P}_{00}\sqrt{\Phi_\trm{m}}}.
\end{equation}
The RHS describes the action of the co-propagating anti-Stokes process.
Similarly, the co-propagating Stokes process drives the $a_1$ mode
\begin{equation}
    \prens{v_1^{-1}\partial_t + \partial_z} a_1\prens{t,z} = -i g_{10} b^* a_0
\end{equation}
with
\begin{equation}
    g_{10} = -\frac{\omega_1}{2}\frac{\int\trm{d}A\,\mathbf{e}_1^*\cdot \delta_\mathbf{u}\perm\cdot\mathbf{u}^*\cdot\mathbf{e}_0 }{\mathcal{P}_{11}\sqrt{\Phi_\trm{m}}}.
\end{equation}
If we flux normalize the optical modes $\mathcal{P}_i = \hbar\omega_i$ and choose the phase of the mode profiles such that $g\equiv g_{01} = g_{10}$, we arrive at
\begin{equation}
    \prens{\mathbf{v}^{-1}\partial_t + \partial_z}\mathbf{a} = -igb\sigma_x\mathbf{a}
    \label{eq:SIPS}
\end{equation}
for $b = b^*$. This is Equation~\ref{eq:dynamics} in the manuscript. By normalizing by flux, the operator on the RHS of Equation~\ref{eq:SIPS} is anti-Hermitian and therefore the dynamics generated by it are unitary. That means photons are scattered between the two modes, conserving the total photon number. This is the Manley-Rowe relations for the processes considered. 

The same formulation is readily adapted to describe other traveling-wave interactions such as electro-optic modulation and non-linear interactions by making a different choice for $\delta\Pi$ or perturbation to the energy $\psi\cdot\delta\Pi\psi/2$~\cite{Yariv1973}.

\subsection{Power orthogonality}
\label{sec:power-orthogonality}

In order to arrive at the diagonalized telegrapher Equation~\ref{eq:telegrapher}, we made use of the fact that the mode profiles $\psi_i$ are power- and energy-orthogonal. Solutions to Equation~\ref{eq:waveguideEigenvalueProblem} simultaneously diagonalize the operators $\hat{z}\times i\sigma_y$ and $\Pi$. Here we show how these orthogonality relations are derived. Similar derivations can be found for optical waveguides in Snyder and Love~\cite{Snyder2012} and piezoelectic waveguides in Auld~\cite{Auld1973}.

The power- and energy-orthogonality relations are closely related to local conservation of energy. First we derive local conservation of energy from Equation~\ref{eq:MEcompact} before deriving from it the orthogonality relations. The operators in Equation~\ref{eq:MEcompact} are symmetric under exchange. For any unit vector $\mathbf{\hat{n}} = \prens{\hat{n},\hat{n}}^\top$, the product $\mathbf{\hat{n}}\cdot\Psi_i\times i\sigma_y\Psi_j$ is invariant under $i\leftrightarrow j$ 
\begin{align}
    \mathbf{\hat{n}}\cdot \Psi_i\times i\sigma_y \Psi_j &= \mathbf{\hat{n}}\cdot\Psi_j\times i\sigma_y \Psi_i \\
        &= \hat{n}\cdot\mathbf{E}_i \times \mathbf{H}_j + \hat{n}\cdot\mathbf{E}_j \times \mathbf{H}_i 
\end{align}
which is manifest when expressed in terms of the fields. Consequently, the divergence takes the symmetric form
\begin{equation}
    -\nabla\cdot\prens{\Psi_i\times i\sigma_y \Psi_j} = \Psi_i\cdot\nabla\times i \sigma_y \Psi_j + \Psi_j\cdot\nabla\times i \sigma_y \Psi_i.
    \label{eq:productRule}
\end{equation}
Substituting in Maxwell's equations we find
\begin{equation}
    -\nabla\cdot\prens{\Psi_i\times i\sigma_y \Psi_j} = \partial_t\prens{\Psi_i\Pi\Psi_j}.
    \label{eq:localConservationOfEnergy}
\end{equation}
so long as $\Pi = \Pi^\top$ and $\partial_t\Pi = 0$. When $i=j$, this is the source-free form of local conservation of energy (multiplied by 2). A similar result follows when $\Pi = \Pi^\dagger$ 
\begin{equation}
    -\nabla\cdot\prens{\Psi_i^*\times i\sigma_y \Psi_j} = \partial_t\prens{\Psi_i^*\Pi\Psi_j}.
    \label{eq:localConservationOfEnergyComplex}
\end{equation}

Power-orthogonality follows from Equation~\ref{eq:localConservationOfEnergyComplex}. Consider two modes $\Psi_i$ and $\Psi_j$ of a waveguide of the form in Equation~\ref{eq:waveguideModes} with $\omega_i = \omega_j$.
In this case, the total time derivative in Eq.~\ref{eq:localConservationOfEnergyComplex} vanishes leaving
\begin{equation}
    \nabla\cdot\prens{\Psi_i^*\times i\sigma_y \Psi_j} = 0.
\end{equation}
Since the modes of the waveguide are confined such that $\psi_i^*\times i\sigma_y \psi_j$ vanishes at the boundary of the $xy$-plane, it follows that
\begin{equation}
    \prens{\beta_j - \beta_i}\int\trm{d}A\,\psi_i^*\times i\sigma_y\psi_j = 0.
\end{equation}
Non-degenerate modes $\beta_i \ne \beta_j$ are power-orthogonal; in a flux-normalized basis, $\mathcal{P}_{ij} = \delta_{ij} \hbar\omega$.

In a similar way we can show that when two modes have equal wavevectors $\beta_i = \beta_j$,
\begin{equation}
    \prens{\omega_j - \omega_i}\int\trm{d}A\, \psi_i^*\Pi\psi_j = 0.
\end{equation}
If $\beta_i = \beta_j$ and $\omega_i \ne \omega_j$, the modes are energy-orthogonal. But what we need to show is that the modes are energy-orthogonal when $\omega_i = \omega_j$ and $\beta_i \ne \beta_j$. To do that we need another constraint on the dynamics.

\subsection{Energy orthogonality}
\label{sec:energy-orthogonality}

We employ time-reversal symmetry to show that two modes, $i$ and $j$, of the same frequency are also energy-orthogonal
\begin{equation}
    \int\trm{d}A\;\psi_i^*\Pi\psi_j = 0.
    \label{eq:app-energy-orthogonality}
\end{equation}
Consider an electromagnetic field of the form
\begin{equation}
    \Psi(t,x,y,z) = \int\trm{d}\omega \tilde{\Psi}(\omega,x,y,z)e^{-i\omega t}.
\end{equation}
Under time-reversal $\mathbf{T}$, each Fourier component of the field state vector becomes
\begin{equation}
    \mathbf{T}\tilde{\Psi} = \sigma_z\tilde{\Psi}^*.
\end{equation}
It follows that $\mathbf{T}$ --- specifically $\sigma_z$ --- flips the sign of the ``power'' operator (LHS of Equation~\ref{eq:MEcompact})
\begin{equation}
    \sigma_z \prens{\nabla\times i\sigma_y} \sigma_z = -\nabla\times i\sigma_y.
\end{equation}
In contrast, so long as $[\sigma_z,\Pi] = 0$, the ``energy'' operator (RHS) does not change sign
\begin{equation}
    \sigma_z \partial_t\Pi \sigma_z = \partial_t\Pi.
\end{equation}
This is what we naively expect: reversing time inverts the power without affecting the energy density. 

When deriving power-orthogonality, the time-dependence of $\Psi_1^*$ and $\Psi_2$ cancels such that the total time derivative $\partial_t\prens{\Psi_1^*\Pi\Psi_2}$ vanishes. We use the relative sign flip of the two operators under $\mathbf{T}$ to preserve these energy terms.

If $[\Pi,\sigma_z] = 0$ and $\Pi = \Pi^*$, then the equations are symmetric under time-reversal. Every harmonic solution $\Psi$ (dropping the tildes) maps to a corresponding time-reversed solution $\mathbf{T}\Psi = \sigma_z\Psi^*$. It follows that 
\begin{equation}
    \prens{\mathbf{T}\Psi_2}^*\cdot\prens{\nabla\times i\sigma_y + \partial_t\Pi}\prens{\mathbf{T}\Psi_1} = 0
\end{equation}
and therefore
\begin{equation}
    \Psi_2\cdot\prens{-\nabla\times i\sigma_y + \partial_t\Pi}\Psi_1^* = 0.
\end{equation}
Subtracting this from 
\begin{equation}
    \Psi_1^*\cdot\prens{\nabla\times i\sigma_y + \partial_t\Pi}\Psi_2 = 0,
\end{equation}
we find
\begin{equation}
    \nabla\cdot\prens{\Psi_1^*\times i\sigma_y \Psi_2} + \Psi_1^*\partial_t\Pi\Psi_2 - \Psi_2\partial_t\Pi\Psi_1^* = 0
\end{equation}
where instead of the total time derivative in Equation~\ref{eq:localConservationOfEnergy}, we have a difference. For waveguide modes, solutions of the form $\psi_i e^{i\beta_i z - i\omega_i t}$, this becomes 
\begin{equation}
    i\prens{\beta_2 - \beta_1}\int\trm{d}A\;\psi_1^*\times i\sigma_y\psi_2 - i\prens{\omega_1 + \omega_2}\int\trm{d}A\; \psi_1^*\Pi\psi_2 = 0
\end{equation}
after integrating over the cross-section.
When $\omega_1 = \omega_2$ and $\beta_1\neq\beta_2$, the modes are power-orthogonal and the first term vanishes, leaving us with the energy-orthogonality relation in Equation~\ref{eq:app-energy-orthogonality}.

Thus at a fixed-frequency when $\Pi = \Pi^\dagger$, $\Pi = \Pi^*$, and $[\sigma_z,\Pi] = 0$, the modes of a waveguide simultaneously diagonalize the operators in Equation~\ref{eq:MEcompact} giving us the independent telegrapher Equations~\ref{eq:telegrapher} and, ultimately, the optomechanically coupled dynamics in Equation~\ref{eq:SIPS}.

\section{Dynamics with loss and dephasing}
\label{app:generalizedDynamics}

The dynamics presented in the text (Equation~\ref{eq:dynamics}) assume the mechanical amplitude $b$ is constant along the waveguide, and the scattering processes are perfectly phase-matched. Here we generalize the model to include loss and dephasing.

\subsection{Mechanical loss}

If we include mechanical loss in our model, the mechanical amplitude \(b\) decays exponentially 
\begin{equation}
    b\prens{z} = b e^{-\gamma z / 2}.
\end{equation}
In the presence of loss, the steady-state solutions become
\begin{equation}
    \mathbf{a}_{\pm}\prens{z} = \mathbf{a}_{\pm}\prens{0} 
        \exp\left\lbrack
            \pm i g b \prens{\frac{1 - e^{-\gamma b z/2}}{\gamma/2}}
        \right\rbrack
        .
    \label{eq:ssSolnWithLoss}
\end{equation}
This yields the same solutions as before (Equation~\ref{eq:ssSoln}) except 
\begin{equation}
    z \rightarrow \frac{2}{\gamma}\prens{1 - e^{-\gamma z / 2}}.
\end{equation}
We use this to define the effective interaction length
\begin{equation}
    L_\trm{eff} = \frac{2}{\gamma}\prens{1 - e^{-\gamma L / 2}}
\end{equation}
which asymptotes to \(2 / \gamma\).

For the SH\(_0\) mode at 440~MHz we measure \(\gamma = -11.7~\trm{dB/mm}\). With this \(\gamma\), \(L_\trm{eff}\) asymptotes to \(742~\um\). In the long-device limit, full conversion \TE{1}$\leftrightarrow$\TE{0} requires
\begin{align}
    P_{\pi/2} &= \frac{\pi^2}{4 \tbmu^2 \, g^2 L_\trm{eff}^2} \\
        &= \frac{36.5~\mu\trm{W}}{\tbmu^2} 
\end{align}
which, given \(\tbmu^2 = - 21.9~\trm{dB}\), is
\begin{equation}
    P_{\pi/2} = 5.65~\trm{mW}
\end{equation}
incident microwave power. This is roughly three orders of magnitude smaller than a bulk AOM but is $10\times$ larger than the efficiency reported by Hinkov \emph{et al.}~\cite{Hinkov1994}.  Improvements to $g$, $\gamma$, and $\tbmu$ are needed to go beyond previous demonstrations. A 10~dB improvement to $\left|\tbmu\right|^2$ is suggested by the -12~dB insertion loss of SH$_0$ waveguide transducers at 2~GHz~\cite{Dahmani2020}.

\subsection{Dephasing and optical bandwidth}
\label{app:mismatched}

Above and in the text we consider phase-matched processes. If the AOM is driven at a different frequency $\Omega$ such that
\begin{align}
    \Delta &\equiv  K + \beta_1\prens{\omega} - \beta_0\prens{\omega + \Omega} \\
        &= 0,
\end{align}
the equations of motion become
\begin{equation}
	\prens{\mathbf{v}^{-1}\partial_t + \partial_z} \mathbf{a} = i g b e^{-\gamma z/2} \sigma_x e^{-i\Delta z\sigma_z}\mathbf{a}.
	\label{eq:EOMcomplete}
\end{equation}
If $\Delta L \gg 1$, the light scattered between modes by the mechanics will destructively interfere and limit the total converted power.

\begin{figure}
    \centering
    \includegraphics[width=0.85\linewidth]{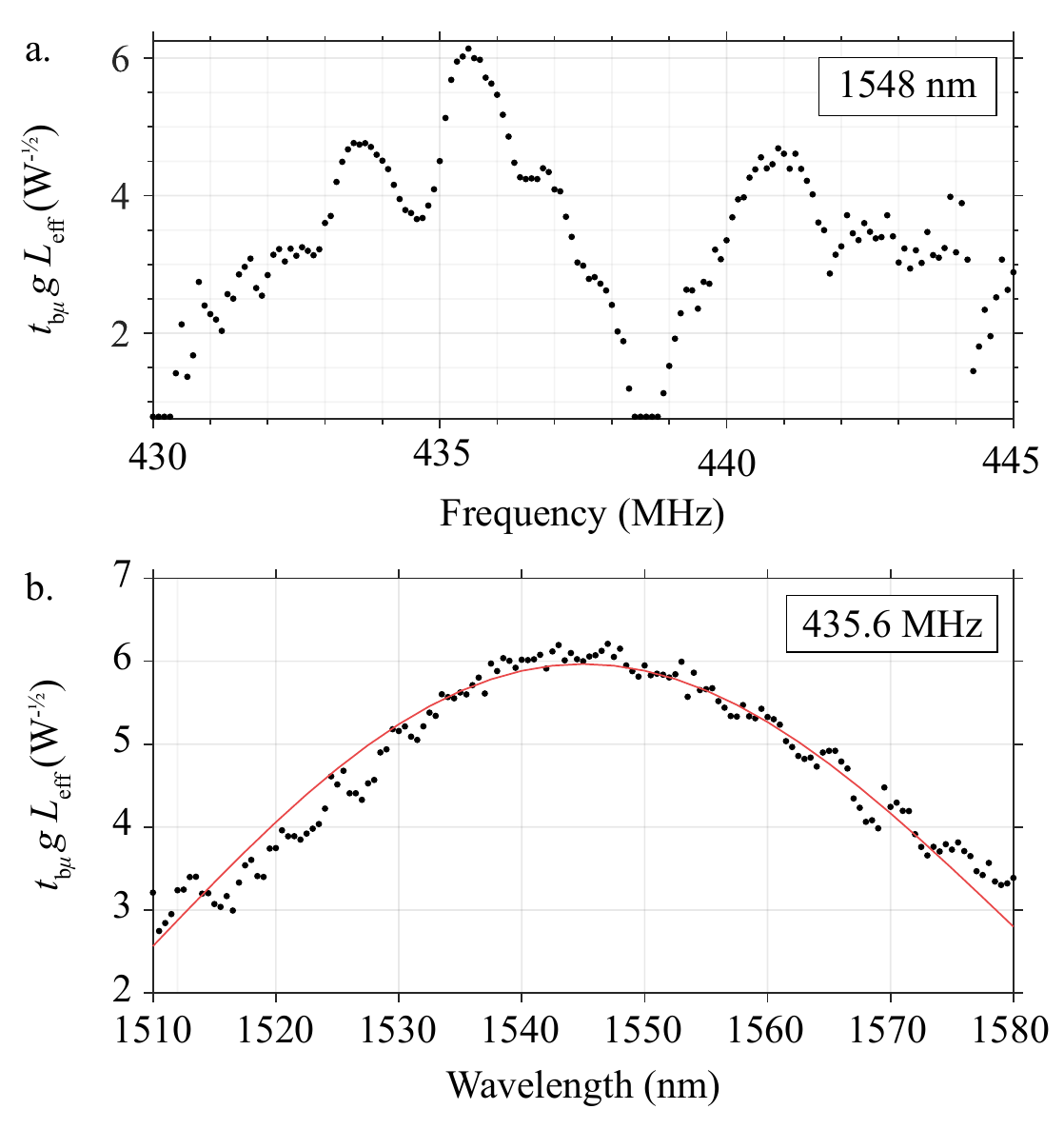}
    \caption{\textbf{Optical and RF bandwidth} The measurements and regression presented in Figure~\ref{fig:heterodyne} are repeated over a range of wavelengths and RF frequencies to determine the bandwidth of the AOM. \textbf{a.} The RF response is largely determined by how the transucer, specifically $\tbmu$, varies with $\Omega$ and so takes a shape similar to the transducer's conductance. \textbf{b.} The optical bandwidth is determined by phase-matching. The red curve is the $g\tbmu L_\trm{eff}$ regressed from a simulated dataset with the measured values for $g$, $\tbmu$, and $\gamma$, and $\Delta n_\trm{g} = 0.175$.}
    \label{fig:bandwidth}
\end{figure}

We repeat the measurements and analysis presented in Figure~\ref{fig:heterodyne}c, varying the RF drive frequency and the optical wavelength away from a phase-matched operating point. The results are plotted in Figure~\ref{fig:bandwidth}. While the RF bandwidth is dictated by the transducer response, the optical bandwidth (Figure~\ref{fig:bandwidth}b) is determined by phase-matching. We simulate a data set from Equation~\ref{eq:EOMcomplete} using the measured values for $g\tbmu L_\trm{eff}$ and an optical group index difference $\Delta n_\trm{g} = 0.175$. Fits to the simulated data are overlaid (red curve) on the measurements. The best-fit $\Delta n_\trm{g}$ is close to the FEM numerical value of $0.125$.

\section{Characterizing the AO Multiplexer}
\label{app:multiplexer}

The performance of the multiplexers are summarized in the manuscript. Here we provide details on their characterization. 

\begin{figure}
    \centering
    \includegraphics[width=0.85\linewidth]{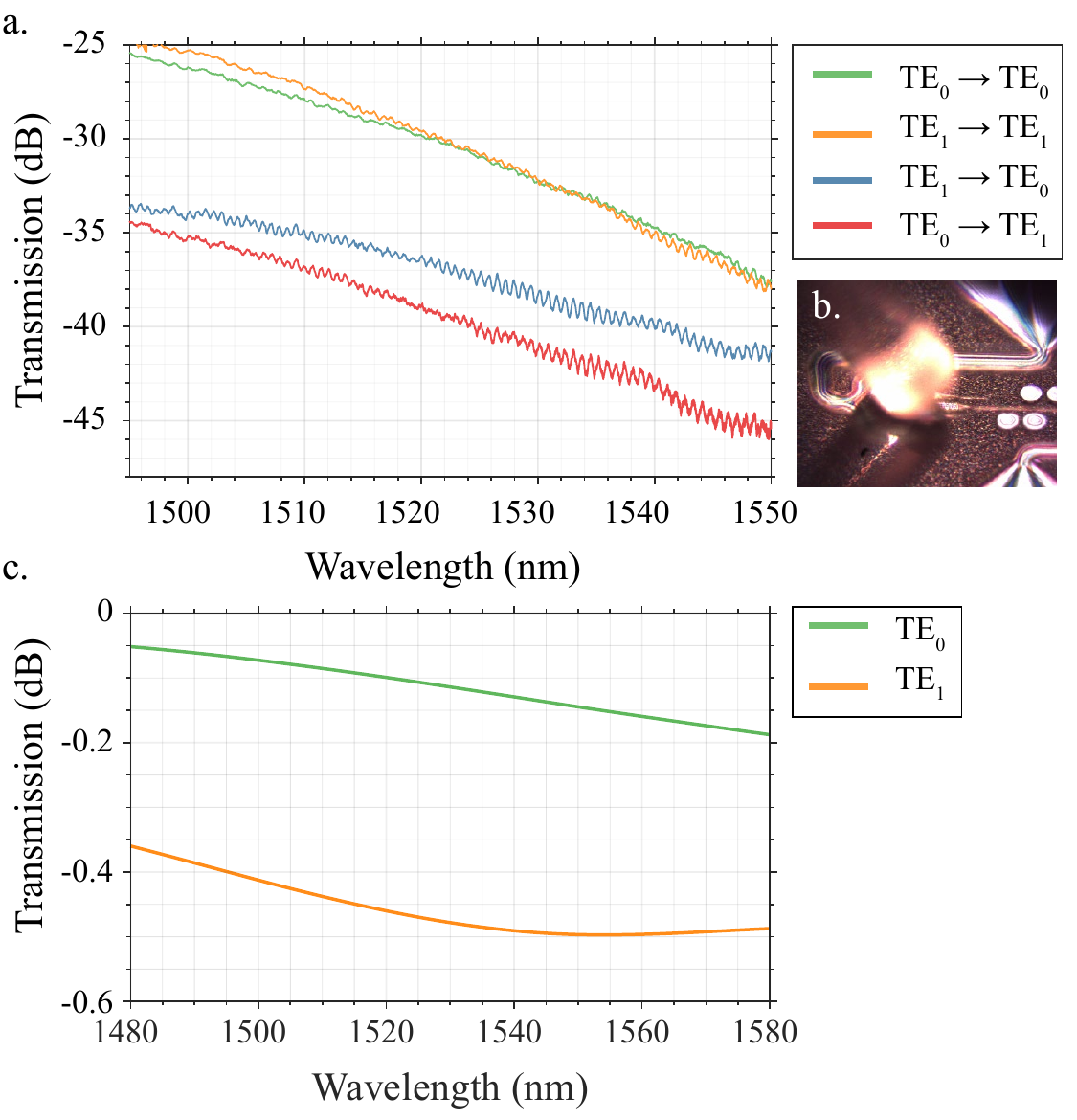}
    \caption{\textbf{Optics} \textbf{a.} FDTD simulations of the insertion loss of both couplers. \textbf{b.} The transmission is measured aligning the fibers to each pair of optical ports.  We see -10~dB crosstalk. The center of the grating response is near 1500~nm. \textbf{c.} Photograph of a fiber coupled to the device. }
    \label{fig:optics}
\end{figure}

\subsection{Optical couplers}
\label{sec:opticalCouplers}

The optical couplers are designed to adiabatically transfer the mode of the coupling waveguide into the AO waveguide. Widths of the two waveguides are chosen by solving for the modes of the adjacent waveguides as discussed in Section~\ref{sec:multiplexer} in the manuscript. The tapered couplers are simulated by FDTD in Lumerical~\cite{Lumerical} and their insertion loss plotted in Figure~\ref{fig:optics}a.

We measure the optical transmission through the device for the four optical paths TE$_{0/1}\rightarrow$TE$_{0/1}$, plotted in Figure~\ref{fig:optics}c. Transmission through the \TE{0} and \TE{1} paths is similar, both exhibiting $-10~$dB crosstalk into the unintended optical mode. This crosstalk limits the isolation of the device. It causes the optical pump to leak through into the signal channel. For example, if light is injected into the \TE{1} port (bottom-left in Figure~\ref{fig:device}), in a perfect device with no crosstalk only photons which absorb a phonon and scatter into the \TE{0} mode leave from the bottom-right. With $-10$~dB crosstalk, 10\% of the pump remaining at the end of the waveguide is sent into the bottom-right port.

The insertion losses for the different paths through the device are unequal which can arise from, \emph{e.g.}, different efficiencies of the two adiabtic optical couplers or variability in the fab. For example, \TE{1}$\rightarrow$\TE{1} can have a higher insertion loss than \TE{1}$\rightarrow$\TE{0}. As a result, if light is fully converted from \TE{1} to \TE{0} by the mechanics, more light can leave the device with the drive on than with the drive off. The \emph{modulation efficiency} as defined in Figure~\ref{fig:heterodyne}d and Figure~\ref{fig:nonreciprocity}b can exceed or not reach 0~dB at the full conversion drive power $P_{\pi/2}$.

\subsection{Piezoelectric transducer efficiency and mechanical propagation loss}
\label{sec:mechanics}

\begin{figure}
    \centering
    \includegraphics[width=0.85\linewidth]{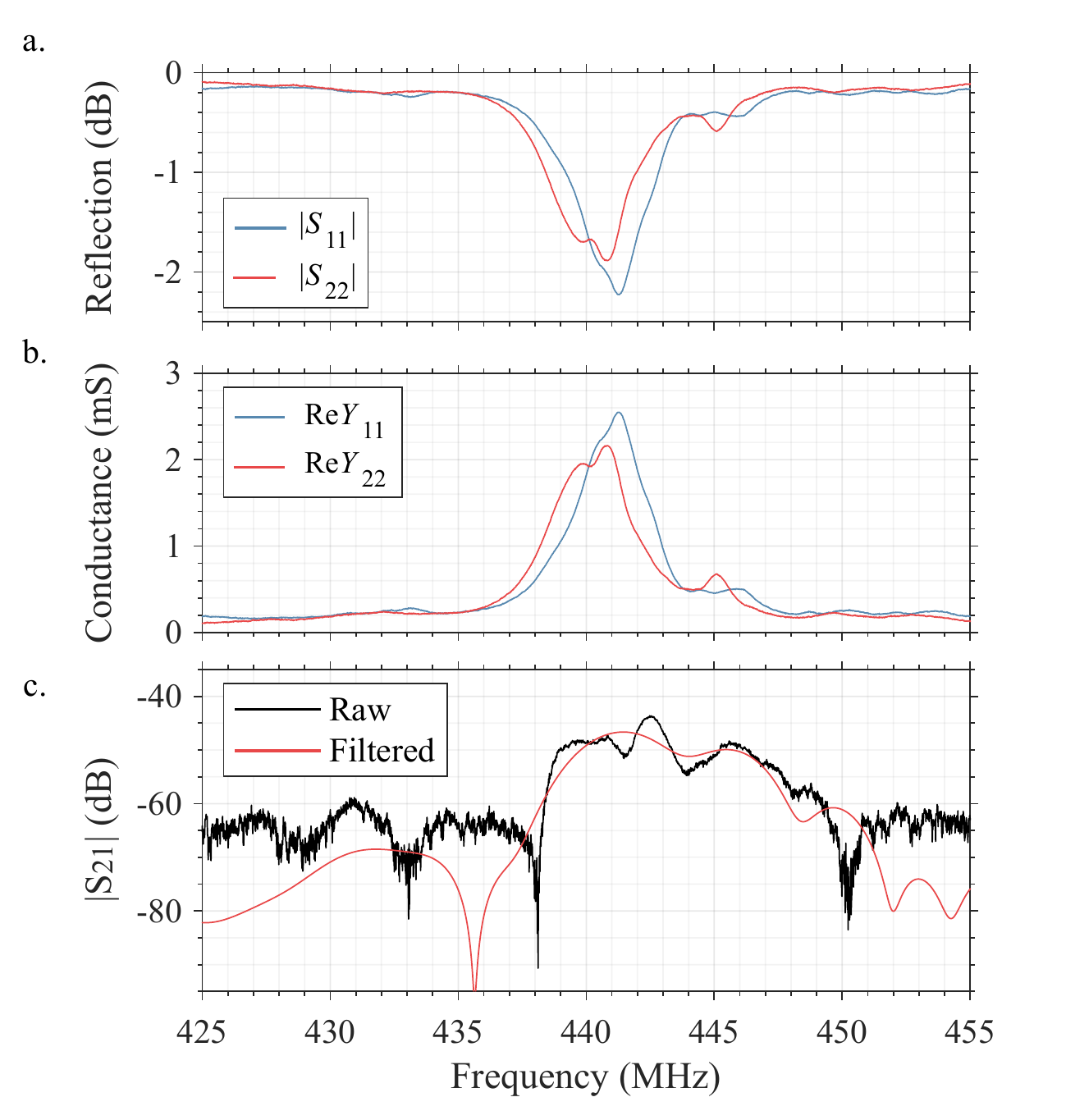}
    \caption{\textbf{Transducer response} The left and right transducer in Figure~\ref{fig:device} are ports 1 and 2, respectively. \textbf{a.} The dip in the reflections near 440~MHz corresponds to the SH$_0$ mode plotted in the bands in Figure~\ref{fig:modes}. \textbf{b.} The conductance reaches 2~mS.\textbf{c.} We overlay the transmission $S_{21}$ filtering out microwave crosstalk and higher-order transits in the device onto the raw data. The filtered response removes ripples from the higher-order transits and is used to extract $\tbmu^2 e^{-\gamma L/2}$. }
    \label{fig:IDTSmatrix}
\end{figure}

The design of the transducer and methods for characterizing its efficiency and the mechanical propagation loss are described in detail in References~\cite{Sarabalis2020sband,Dahmani2020}. The S-matrix of the device in Figure~\ref{fig:device} is measured on a calibrated probe station. From the reflection measurements $S_{ii}$ where $i = 1$ for the left transducer and $i = 2$ for the right, we see that the response of the transducer is repeatable. The reflections (Figure~\ref{fig:IDTSmatrix}a) reach -2~dB for the SH$_0$ response corresponding to a -4.3~dB loss from impedance mismatch between the $50~\Omega$ transmission line and transducer. 

\begin{figure}
    \centering
    \includegraphics[width=\linewidth]{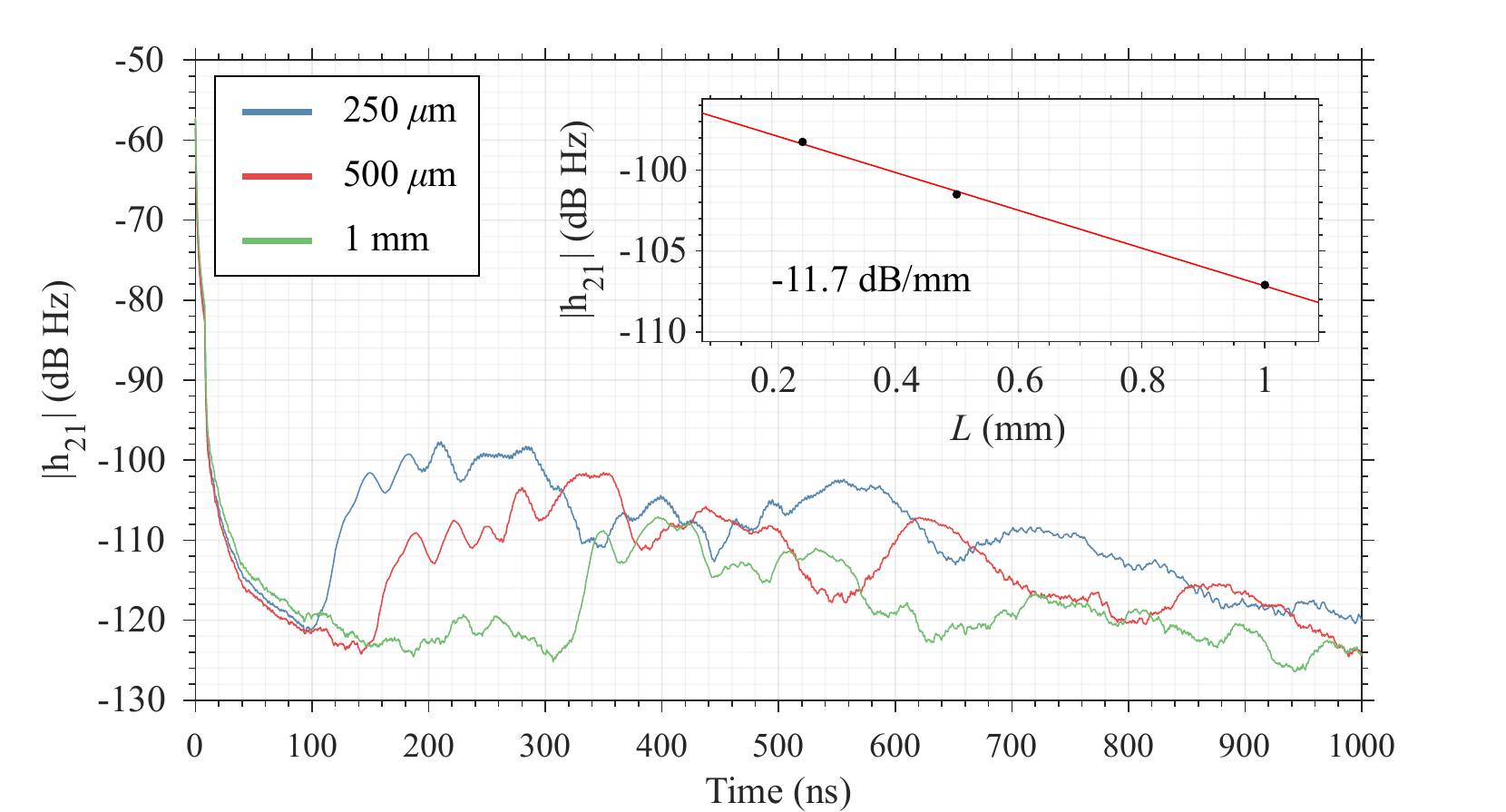}
    \caption{\textbf{Mechanical propagation loss} We extract the propagation loss from a length sweep of the waveguide by fitting a line to the magnitude of the impulse response. We fit an exponential to the peak response yielding $\gamma = -11.7~\trm{dB/mm}$ (inset). }
    \label{fig:IDTLengthSweep}
\end{figure}

The single-port response is not enough to determine the transducer's efficiency. There are other loss mechanisms in addition to impedance mismatch which reduce the transducer's efficiency. Of the power emitted into the waveguide, nearly all of it is in the SH$_0$ mode but, as discussed in detail in Reference~\cite{Dahmani2020}, resonances in the transducer lead to large material damping losses. In order to determine the transmission coefficient $\tbmu$ from microwaves in the line to phonons in the waveguide, we need to measure the two-port response $S_{21}$ and filter out contributions from triple-transit~\emph{etc.}~\cite{Sarabalis2020sband}. In Figure~\ref{fig:IDTSmatrix}c, we plot the raw $S_{21}$ as well as the single-transit response filtered in the time-domain. This filtered signal is equal to $\tbmu^2 e^{-\gamma L/2}$, assuming symmetric coefficients for the two transducers. The mechanical losses are independently determined by measuring how the transmission varies with the length of the waveguide $L$. The impulse response $h_{21}$, \emph{i.e.} the Fourier transform of $S_{21}$, is plotted in Figure~\ref{fig:IDTLengthSweep} where the peaks are fit (inset) for the propagation loss $\gamma$.

\section{RF power-handling}
\label{sec:power-handling}

\begin{figure}
    \centering
    \includegraphics[width=\linewidth]{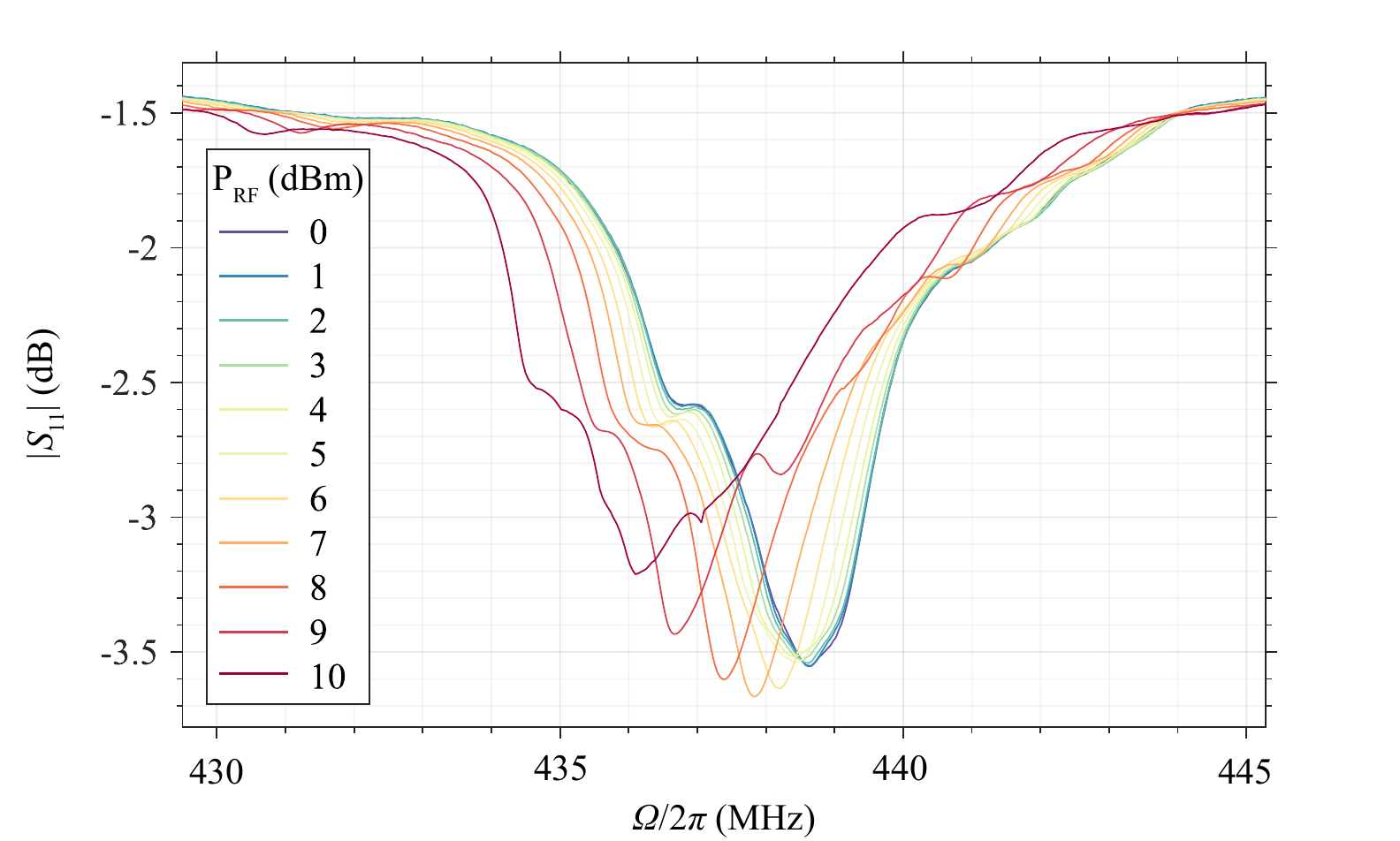}
    \caption{\textbf{Power-handling of the Transducer} Microwave reflections $S_{11}$ form the transducer show a center frequency shift and bi-stabilities as the drive power $P_\trm{RF}$ is increased to $10~\trm{dBm}$.}
    \label{fig:power-handling}
\end{figure}

The maximum conversion efficiency observed $\eta_\trm{max} = 18\%$ is limited by the microwave power-handling of the transducer and the AO multiplexer. In Figure~\ref{fig:power-handling}, we plot reflections from the transducer $S_{11}$ measured on a vector network analyzer as the power is increased to $10~\trm{dBm}$ at which $\eta_\trm{max}$ was observed. As the power is increased, the center frequency of the IDT decreases and, for fixed-frequency drives, can cause drops above 1~dB. After removing the 1.2~dB round-trip loss in the cable, this change in $S_{11}$ amounts to a $-3~\trm{dB}$ decrease in the power delivered to the device. This power-dependent frequency shift causes the curves in Figure~\ref{fig:heterodyne}c to deviate from a sinusoid, and so we restrict our fits for $g\tbmu$ to the low-power portion of the dataset. 

Not only do we see frequency shifts, at $10~\trm{dBm}$ discontinuities appear in $S_{11}(\Omega)$, evidence of bi-stability arising from, \emph{e.g.}, Duffing nonlinearities. 

At 13~dBm drive power, the tethers in the multiplexer broke. The AO and coupling waveguides separated from one another, destroying the optical couplers.

\section{Inferring \(g\)}
\label{app:inferring_g}

In our heterodyne measurements described in Section~\ref{sec:characterization}, both the pump depletion and converted signal powers are measured, and both can be used to infer $g$. In the low-power limit, the pump depletion provides a better measure of the $P_{\pi/2}$. Given a pump with initial amplitude $a_0$, the flux in the pump varies up to $\mathcal{O}\prens{\beta^5}$ as 
\begin{equation} 
    \left|a_0\right|^2 \cos^2\prens{g\tbmu \beta L} \approx \left|a_0\right|^2 \prens{ 1  - \left| g \tbmu \beta L\right|^2 + \frac{1}{3}\left| g \tbmu \beta L\right|^4}.
\end{equation}
while the signal power varies as
\begin{equation} 
    \left|a_0\right|^2 \sin^2\prens{g\tbmu \beta L} \approx \left|a_0\right|^2 \prens{ \left| g \tbmu \beta L\right|^2 + \frac{1}{3}\left| g \tbmu \beta L\right|^4}.
\end{equation}
Two terms in the series are needed to fit $g\left|\tbmu\right| L$ without independently measuring $\left|a_0\right|^2$. This is satisfied for the pump to second-order in $\beta$ but to fourth order for the signal, making pump depletion a more sensitive measure of the efficiency in the low-power limit.

The drives used for the dataset in Figure~\ref{fig:heterodyne} were high enough for independent regressions on the pump and converted signal to give comparable values for $P_{\pi/2}$.

\end{document}